\documentclass[twocolumn,showpacs,preprintnumbers,prb,fleqn,floatfix]{revtex4}

\usepackage{graphicx}
\usepackage{dcolumn}
\usepackage{bm}
\usepackage{amsmath,mathrsfs}
\renewcommand{\Im}{i}

\newcommand{\Ef}{E_{\rm F}}
\newcommand{\D}{\Delta}

\newcommand{\wc}{\omega_{\rm c}}
\newcommand{\Oc}{\Omega_{\rm c}}
\newcommand{\mc}{m_{\rm c}}
\newcommand{\Mc}{M_{\rm c}}
\newcommand{\mh}{m_z}
\newcommand{\Mh}{M_z}
\newcommand{\mub}{\mu_{\rm B}}
\newcommand{\me}{m_{\rm e}}

\newcommand{\kp}{$\bm{k}\cdot\bm{p}$ }

\begin{document}
\title{ Angle-resolved Landau spectrum of electrons and holes in bismuth}
\author{Zengwei Zhu$^{1}$, Beno\^{\i}t Fauqu\'e$^{1}$, Yuki Fuseya$^{2}$ and Kamran Behnia$^{1}$}
\affiliation{(1) Labotoire de Physique Et d'Etude des Mat\'eriaux (UPMC-CNRS)\\
ESPCI, 10 Rue Vauquelin, 75005 Paris, France\\
(2) Department of Materials Engineering Science\\
Osaka University, Toyonaka, Osaka 560-8531, Japan}

\date {August 22, 2011}

\begin{abstract}
In elemental bismuth, emptying the low-index Landau levels is accompanied by giant Nernst quantum oscillations. The Nernst response sharply peaks each time a Landau level intersects the chemical potential. By studying the evolution of these peaks when the field rotates in three perpendicular planes defined by three high-symmetry axes, we have mapped the angle-resolved Landau spectrum of the system up to 12 T. A theoretical model treating electrons at L point with an extended Dirac Hamiltonian is confronted with the experimentally-resolved spectrum.  We obtain a set of theoretical parameters yielding a good but imperfect agreement between theory and experiment for all orientations of the magnetic field in space. The results confirm the relevance of the Dirac spectrum to the electron pockets and settle the longstanding uncertainty about the magnitude of the g-factor for holes. According to our analysis, a magnetic field exceeding 2.5 T applied along the bisectrix axis puts all carriers of the three electron pockets in their lowest($j=0$) spin-polarized Landau level. On top of this complex angle-dependent spectrum, experiment detects additional and unexpected Nernst peaks of unidentified origin.

\end{abstract}


\maketitle

\section{Introduction}

Bismuth crystallizes in a rhombohedral structure. This is an intriguing alternative to the more symmetric cubic symmetry, favored by the balance in electron energetics\cite{peierls,abrikosov}. The departure from higher symmetry has made it a semi-metal with a significant role in the history of metals physics (for reviews see\cite{dresselhaus,edelman,issi}). In particular, it was in bismuth that a Fermi surface was experimentally determined for the first time seventy years ago\cite{shoenberg}. This historical priority is not a consequence of intrinsic simplicity. The structure of Fermi surface in bismuth is quite unlike the sphere seen in the opening chapters of a standard condensed-matter textbook on the free electron gas. On the other hand, and in spite of its complexity, this Fermi surface is one of the easiest to detect by experiment. Indeed, because of the lightness and high mobility of the carriers, quantum oscillations in bismuth are observable at relatively high temperatures and low magnetic fields.

\begin{figure}
\resizebox{!}{0.18\textwidth}{\includegraphics[bb=0 0 565 223]{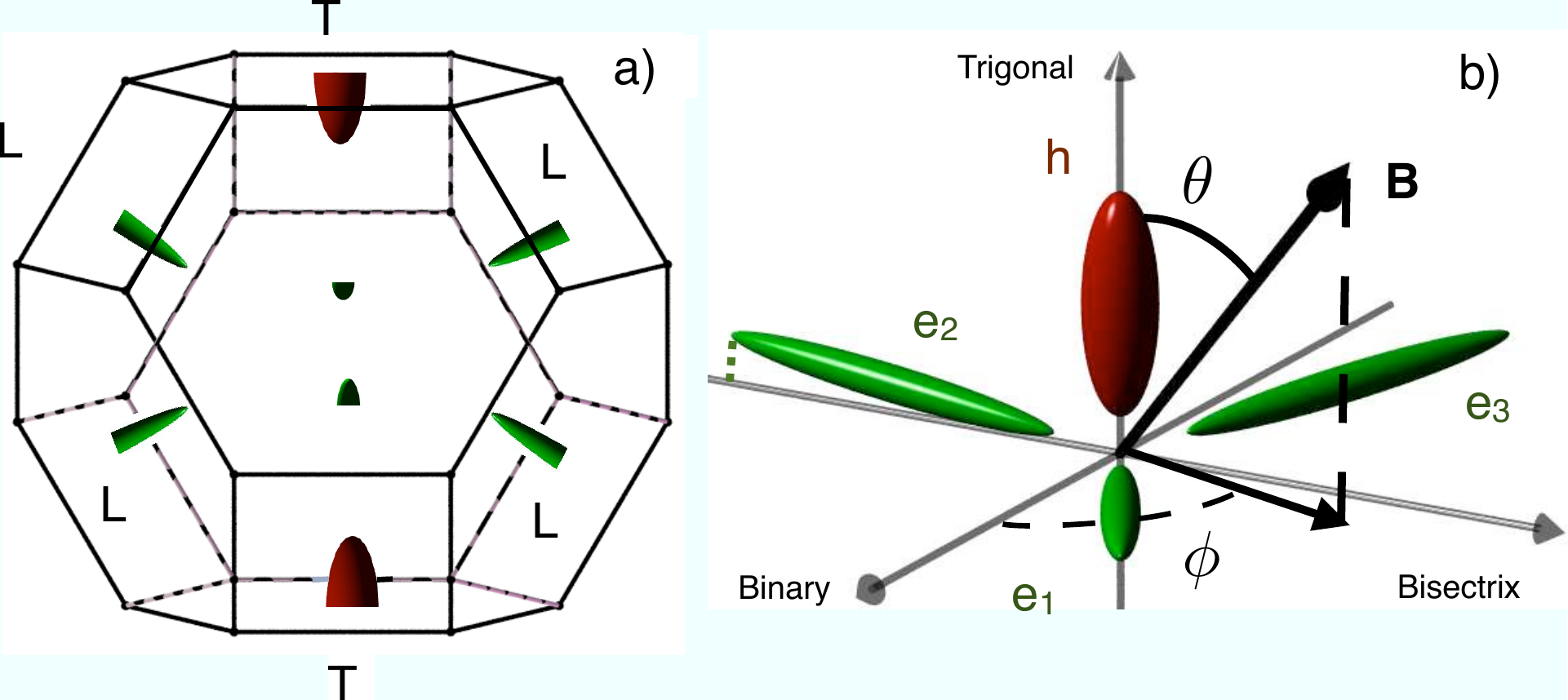}}\caption{ a) Sketch of the Brillouin zone and the Fermi surface of bismuth. The size of the Fermi surface is enhanced to make it visible. It consists of one hole pocket at the T point and three electron pockets at the L points of the Brillouin zone. b) In our experiment, $\theta$ is the angle between the magnetic field and the trigonal axis, $\phi$  is the angle between the binary axis and the projection of the magnetic field in the equatorial plane (defined by the binary and bisectrix axes) The three electron pockets are noted e$_1$, e$_2$ and e$_3$ as seen in the figure according to their position to the angle $\phi$. }
\end{figure}
During several decades, numerous investigations of the de Haas-van Alphen and Shubnikov- de Haas effects were carried on bismuth in order to pin down the fine details of the Fermi surface\cite{steele,smith,brandt,bhargava,brown1,mangez,hiruma1,hiruma2,michenaud,yang,bompadre}. The well-established structure is sketched in Fig.1. It consists of one hole ellipsoid aligned perpendicular to the plane in which lay obliquely three slightly tilted\cite{brown2} electron ellipsoids. Thanks to these intensive studies, the size of each ellipsoid and the tilt angle of the electron ellipsoids are known with a remarkable precision\cite{bhargava}.

In parallel to these experiments, a number of band calculations have been reported\cite{golin,gonze,liu,shick}. However, the relevant energy gap of the system is in the range of $10-20 meV$ and therefore beneath the typical energy resolution of \emph{ab initio} calculations. The most successful band picture has been a tight-binding model using semi-empirical parameters\cite{liu}. The results of this model are in rather good agreement with the details of the Fermi surface according to the experiment.

Recently, there has been a renewal of interest in the electronic spectrum of bismuth, triggered by the relatively easy access it provides to the quantum limit\cite{bompadre,behnia2,behnia3} and the presence of Dirac fermions\cite{luli,fuseya}. The quantum limit is attained when the magnetic field is strong enough to confine electrons to their lowest Landau level. For ordinary bulk metals, the size of the required magnetic field is well beyond the limits of current technology. In a low-density semi-metal such as bismuth, on the other hand, a modest magnetic field applied along the high-symmetry axis known as the trigonal axis allows one to attain this limit. High-field experiments reporting peaks in the Nernst coefficient\cite{behnia3,fauque,yang2}, jumps in magnetization\cite{luli} and Hall plateaus\cite{fauque2} raised the issue of collective effects in a three-dimensional electron gas at high magnetic fields\cite{halperin,tesanovic,macdonald}.

In the context of these unexpected experimental findings, the high-field phase diagram of bismuth became subject to a theoretical reinvestigation\cite{sharlai,alicea}. The Landau level spectrum of a Fermi liquid is determined by the detailed topology of the Fermi surface. In bismuth, this spectrum becomes complex in the vicinity of the quantum limit for several reasons. To preserve charge neutrality, any modification in density of one type of carriers, caused by emptying a Landau level, should be compensated with an equal change in the density of the carriers of the other sign. The chemical potential shifts as the magnetic field is scanned and quantum oscillations are no longer a periodical function of $B^{-1}$. Moreover, the large spin-orbit coupling leads to an angle-dependent Zeeman energy. For both electrons and hole, the anisotropy of the ``spin mass'' is different from the anisotropy of the cyclotron mass. As early as 1964, Smith et \emph{al.} investigated this feature\cite{smith}.

Dirac fermions provide a major motivation to revisit the Landau spectrum of bismuth. It has been known for a long-time that the energy-momentum dispersion of electrons at the L-point is not parabolic and that Dirac Hamiltonian is appropriate to treat them\cite{Cohen1960,Lax1960,cohen,wolff}. Dirac fermions have been identified as the source of the enhanced diamagnetism in bismuth\cite{fukuyama}, and may generate an unusual weak-field Hall conductivity\cite{fuseya}. During the past decades, several theoretical models of three-dimensional Dirac fermions have been proposed and confronted with the experimental data\cite{takaoka,McClure1976,vecchi,hiruma2}, but in contrast to the well-established topology of the Fermi surface\cite{bhargava}, the details of the relevant Hamiltonian has not been pinned down.

During the last few years,  bismuth has attracted attention from other angles of research. In particular, infrared conductivity measurements have detected a collective electron-plasmon mode (a ``plasmaron")\cite{tediosi} and the insulating-like magnetoresistance has been a subject of debate and investigation\cite{du,kopelevich}.

In this paper, we report on a study of angle-dependent Nernst effect in bismuth, which maps the evolution of Landau levels of electrons and holes in the whole solid angle up to 12 T. The data are compared to the results of a theoretical calculation based on a model with an extended Dirac Hamiltonian. The comparison between theory and experiment narrows down the choice of parameters for this model.  Two previous reports have mapped the angle-dependent Landau spectrum of bismuth in a restricted angular window. Smith and co-workers\cite{smith} used Shubnikov- de Haas measurements to determine the Landau spectrum in (trigonal, binary) and (trigonal, bisectrix) planes. This pioneer work established the bulk of electronic properties, but left unsettled the precise magnitude of the hole Zeeman energy. Kajimura and co-workers\cite{kajimura} used giant oscillations of ultrasonic attenuation to obtain a high-resolution map of the Landau spectrum near the trigonal axis. Our study, continuing this line of exploration,  uses giant quantum oscillations of the Nernst coefficient to obtain a high-resolution map in the whole solid angle. The verification of the non-interacting spectrum in bismuth provides a basis to find an explanation for the additional features (i.e. Nernst peaks\cite{behnia3,fauque,yang2}) which are not expected in this picture.

Among the details of the one-particle picture coming out of our investigation, let us highlight two. First, our results definitely settle the uncertainty on the magnitude and anisotropy of the Zeeman energy for holes. Second, when the field is along the bisectrix, the spectrum of the three electron pockets is very close to a perfect Dirac spectrum where the cyclotron and Zeeman energies are indistinguishable. In other words, spin and orbital angular momenta are described by a single quantum number. In this configuration, a magnetic field as low as 2.5 T is enough to attain the quantum limit of all electron pockets to put them in their lowest spin-polarized Landau sub-level. In other words, all electron-like carriers become spin polarized, residing in their lowest ($j=0$; where $j$ is the total angular momentum) Landau level.

In addition to the case of bulk semi-metallic bismuth, the results reported here may have implications for the analysis of the electronic properties of topological insulators\cite{hasan}. In particular, in the case of Bi$_{1-x}$Sb$_{x}$ alloys\cite{teo}, the electronic spectrum of bulk bismuth and bulk antimony is the starting point to analyze the surface states\cite{hofmann} of these systems.

\section{Theoretical treatment of the spectrum at $L$ point}
\begin{figure}
\resizebox{!}{0.25\textwidth}{\includegraphics[bb=0 0 612 298]{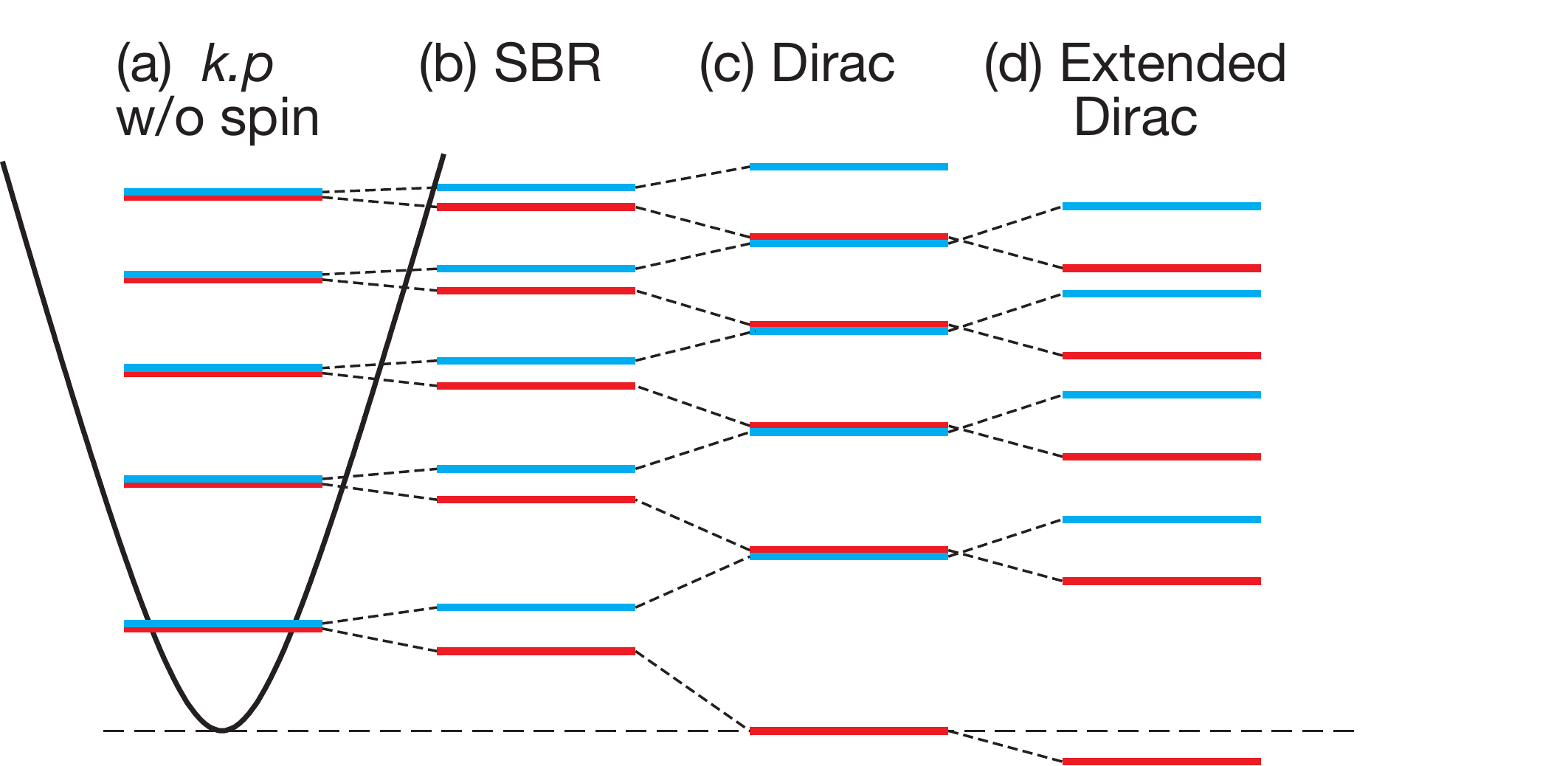}}\caption{Evolution of the bismuth model. The horizontal lines indicates the position of Landau levels with opposite spins.}
\label{models}
\end{figure}

	The essential properties of electrons at $L$ point are the following:
	(1) The Fermi surface is ellipsoidal and tilted.
	(2) The electronic dispersion is non-parabolic.
	(3) The  Landau levels display a slight spin splitting.
	The first property was already realized at the early stage of investigation\cite{shoenberg}.
	The second one, the non-parabolicity of the dispersion, can be naturally obtained on the basis of the \kp perturbation theory.
	The \kp method was first applied to bismuth by Cohen and Blount\cite{Cohen1960}. Soon afterwards, Lax and co-workers used this model to analyze their experimental data\cite{Lax1960}.
	Because of the narrow gap character of $L$ point, the number of bands needed is small and a model with coupled two bands gives a very good agreement with experiment.
	The \kp theory for a two-band model \emph{without spin} gives the energy under a magnetic field in the form:
	$
		E= \pm \left[ \D^2 + 2\D \left\{ \left( n + 1/2 \right) \hbar \omega_{\rm c} + \hbar^2 k_z^2 /2\mh\right\}\right]^{1/2},
	$
	where $\omega_{\rm c}$ is the cyclotron frequency and $\mh$ is the effective mass along the direction of the magnetic field.
	The third property, the spin splitting\cite{Boyle1960,Kunzler1962,smith}, can be taken into account if we introduce the spins with an effective g-factor (Fig. \ref{models} (b)).
	The energy now becomes
$
		E= \pm \left[ \D^2 + 2\D \left\{ \left( n + 1/2 \right) \hbar \omega_{\rm c} + \hbar^2 k_z^2 /2\mh \pm g_{\rm eff}\mub B/2\right\}\right]^{1/2},
$
	where $\mub = |e| \hbar / 2\me c$ is the Bohr magneton.
	In spite of its very simple form, this two-band model, which we will call the Smith-Baraff-Rowell (SBR) model, gives a very good account of the experiment, in particular when the magnetic field  is oriented close to the trigonal axis\cite{Lax1960,smith,sharlai}.
	However,  there is a serious intrinsic problem in the SBR model. When the term $-g_{\rm eff}\mub B/2$ is large and negative, the energy becomes imaginary. As a consequence, it is difficult or impossible to use this model when the field is large and oriented far off the trigonal axis.

	In the original Cohen-Blount theory, the \kp theory was derived including the spin-orbit coupling (not fully contained in the SBR model), which is extremely large in bismuth ($E_{\rm SO}\sim 1.5$eV\cite{gonze}).
	Wolff\cite{wolff} found that the Cohen-Blount theory is essentially identical to the Dirac theory in the relativistic quantum mechanics. This finding led to the birth of the Dirac electron in solids.
	The two band \kp model with spin-orbit coupling (Dirac model\cite{note1}) gives the energy\cite{note2} in terms of a quantum number $j=n+1/2+s$ as:
	$
		E= \pm \left[ \D^2 + 2\D\left\{ j \hbar \omega_{\rm c} + \hbar^2 k_z^2 /2\mh\right\}\right]^{1/2}.
	$
	With this Dirac model, the spin-orbit coupling is taken into account and the problem of imaginary energy is removed, so that it can be safely used for all field directions.
	On the other hand, in this model, the spin splitting is exactly the same as the Landau level splitting, and a double degeneracy occurs for all Landau levels except for the lowest Landau level $j= 0$(Fig. \ref{models} (c)). This is not in agreement with the  experimental observation of spin splitting.

	To repair this discrepancy and to understand the spin splitting, Baraff extended the Dirac model by considering the effect of other bands outside the closely coupled two bands based on the perturbation theory\cite{Baraff1965}.
	As a result, he showed that it is possible to produce a difference between the orbital level splitting and the spin splitting (Fig. \ref{models} (d)).
	Although the Baraff's model is complex to use for the interpretation of experimental results, Dresselhaus and co-workers\cite{Maltz1970,dresselhaus,Vecchi1974,vecchi} have succeeded to simplify the Baraff's model and to obtain theoretical results in good agreement with experiments at least when the magnetic field is oriented along the binary and bisectrix axes.
	In the simplified model, the energy is given by
	$
		E= \pm \left[ \D^2 + 2\D\left\{ j \hbar \omega_{\rm c} + \hbar^2 k_z^2 /2m_z\right\}\right]^{1/2} \pm g'\mub B/2,
	$
	where the additional g-factor expresses the effect of outside bands.
	With this extended Dirac model, all three properties above are taken into account, and no imaginary energy appears in the entire solid angle.
	The number of parameters is the same as the SBR model and it is as easy to handle as the SBR model.
	Consequently, the extended Dirac model will be the most appropriate model for the present purpose.

	It should be noted here that the energy band model of bismuth has been the subject of an alternative approach, which starts with  a simple cubic lattice without spin-orbit coupling. The Hamiltonian for bismuth is then expanded to first order with (i) the distortion of the lattice and then with (ii) the spin-orbit coupling, and (iii) the distance in $k$ space from the $L$ point.
	This approach was first developed by Abrikosov and Falkovsky\cite{abrikosov}, and established by McClure\cite{McClure1976}.
	One important consequence of this approach is that the dispersion contains in addition to a term linear in $k$ term a quadratic term. This is the so-called non-ellipsoidal non-parabolic dispersion equivalent to the form found by Cohen\cite{cohen,note2}.
	It is also the characteristic of this model that the Hamiltonian is directly related to the relative displacements of the atoms, the rhombohedral deformation, and the spin-orbit couplings, so that we can predict the effect of substitution of other elements or the effect of pressure.
	The obtained dispersion is similar to the Dirac model except for the presence of quadratic (non-ellipsoidal) term.
	However,  there is no compelling evidence to prefer the non-ellipsoidal model over the ellipsoidal model\cite{dresselhaus}, in spite of various attempts to use the non-ellipsoidal model for the analysis of experimental data\cite{Weiner1962,Kao1963}.
	Moreover, it is not clearly understood whether this non-ellipsoidal model can reproduce the experimentally resolved spin splitting.
\section{The Model}
\begin{table}
\caption{\label{table1} All theoretical parameters for the electron mass $m_i$ and the additional g-factor $g'_i$, and for the orbital- and spin mass of holes, which give the best fitting results with experiments. The tilt angle of the electron pocket is $6.2^\circ$.}
\begin{ruledtabular}
\begin{tabular}{ccccc}
Electrons & $m_1$ & $m_2$ & $m_3$ & $m_4$\\
\hline
 & 0.00124 & 0.257 & 0.00585 & -0.0277\\
\hline \hline
 & $g'_1$ & $g'_2$ & $g'_3$ & $g'_4$\\
 \hline
 & -7.26 & 24.0 & -7.92 & 9.20
 \end{tabular}
 \label{electron mass}
\end{ruledtabular}
\vspace{1cm}
\begin{ruledtabular}
\begin{tabular}{lcc}
 Holes & $M_1 = M_2$ & $M_3$ \\
 \hline
 orbital mass & 0.0678 & 0.721\\
 spin mass & 0.0319 & 200
\end{tabular}
\end{ruledtabular}
\vspace{1cm}
\begin{ruledtabular}
\begin{tabular}{lcc}
 gap at $L$ point ($2\D$)& 15.3 meV \\
electron-hole hybridization ($E_0$) & 38.5 meV
\end{tabular}
\end{ruledtabular}
\end{table}
\begin{table}
\caption{\label{table2} Various physical values along the principal axes. $\mc$ and $\Mc$ refer the cyclotron mass of electrons and holes.
$m_z$ and $M_z$ are the band mass of electrons and holes along the field orientation.  They are all normalized by the bare mass of electron, $\me$.}
\begin{ruledtabular}
\begin{tabular}{lccc}
  & $\bm{B}\parallel$ Binary &  $\bm{B}\parallel$ Bisectrix &  $\bm{B}\parallel$ Trigonal \\
 \hline
$\mc^{\rm e2}$ & 0.0272 & 0.00189 & 0.0125\\
$\mc^{\rm e1, e3}$ & 0.00218 & 0.00375 & 0.0125\\
$\mh^{\rm e2}$ & 0.00124 & 0.257 & 0.00585\\
$\mh^{\rm e1, e3}$ & 0.193 & 0.0653 & 0.00585\\
$g^{\rm e2}$ & 73.5 & 1060 & 159\\
$g^{\rm e1, e3}$ & 917 & 533 & 159\\
$g'^{\rm e2} $ & -7.26 & 24.0 & -7.92\\
$g'^{\rm e1, e3} $ &16.2 & 0.545 & -7.92\\
$1+(\mc g' )^{\rm e2}/2$ & 0.90 & 1.02 &  0.950\\
$1+(\mc g')^{\rm e1, e3}/2$ & 1.01 &  1.00 &  0.950\\
\hline
$\Mc$ & 0.221 & 0.221 & 0.0678\\
$\Mh$ & 0.0678 & 0.0678 & 0.721\\
$G$ & 0.791 & 0.791 & 62.6\\
$E_{\rm Z} /\hbar \wc$ & 0.0875 & 0.0875 & 2.12
\end{tabular}
\end{ruledtabular}
\end{table}
	The effective Hamiltonian for electrons in bismuth at $L$ point in the Brillouin zone can be described in terms of the two-band model (written as a $4\times4$ matrix including the spin degrees of freedom).
	The Dirac Hamiltonian is given by\cite{wolff,vecchi}
	\begin{align}
		\mathscr{H}_0=
		\begin{pmatrix}
			\D  & \Im \sqrt{\D} \bm{K}\cdot \bm{\sigma} \\
			-\Im \sqrt{\D} \bm{K}\cdot \bm{\sigma} & -\D
		\end{pmatrix},
	\end{align}
	where $\bm{\sigma}$ is the Pauli matrix and the $2\times2$ unit matrix is omitted.
	When the direction of the magnetic field is chosen as direction $z$, and the directions $x$ and $y$ are chosen to be perpendicular to each other, the (quasi) wave vector $\bm{K}$ is related to the crystal wave vector by
	\begin{align}
		K_x \pm \Im K_y &= \frac{\hbar(\tilde{k}_x \pm \Im \tilde{k}_y)}{\sqrt{\mc}}, \\
		K_z& = \frac{\hbar k_z}{\sqrt{\mh}},
	\end{align}
	with the cyclotron mass $\mc$, the longitudinal mass $\mh$, $\tilde{k}_x = k_x + (eB/2c) y$ and $\tilde{k}_y = k_y - (eB/2c) x$, i.e., the symmetric gauge.
	The wave vector $\bm{K}$ satisfies the commutation rule:
	\begin{align}
		\bm{K}\times \bm{K} = \frac{\Im \hbar e \bm{B}}{\mc c}.
	\end{align}
	In Wolff's version of Dirac theory, the eigenvalues of this Hamiltonian can be obtained by considering its squared equation,
	\begin{align}
		\mathscr{H}_0^2 \psi &=
		\begin{pmatrix}
			\D^2 + \D (\bm{K}\cdot \bm{\sigma})^2 & 0 \\
			0 & \D^2 + \D (\bm{K}\cdot \bm{\sigma})^2
		\end{pmatrix}
		\psi \nonumber \\
		&= E^2 \psi.
\end{align}
	The quantity $(\bm{K}\cdot \bm{\sigma})^2$ is written as
	\begin{align}
		(\bm{K}\cdot \bm{\sigma})^2 = K^2 - \frac{\hbar e}{\mc c}\bm{\sigma}\cdot \bm{B},
	\end{align}
	by using the commutation rule.
	Then, the squared wave equation becomes
	\begin{align}
		\begin{pmatrix}
		 	\mathscr{H}^* & 0\\
			0 & \mathscr{H}^*
		\end{pmatrix}
		\psi
		=\frac{E^2 - \D^2}{2\D} \psi,
		\label{9}
\end{align}
	where
	\begin{align}
		\mathscr{H}^* = \frac{K^2}{2} + g \mu_{\rm B} \bm{s}\cdot \bm{B}.
\end{align}
	Here we introduced the effective g-factor
	\begin{align}
		g = 2\frac{\me}{\mc},
\end{align}
	with the spin operator $\bm{s}=\bm{\sigma}/2$.
	%
	%
	If we introduce a ``non-relativistic" energy $E'=E-\D$ ($\ll \D$), the right hand side of Eq. (\ref{9}) is approximated as
	\begin{align}
		\frac{E^2 - \D^2}{2\D } \psi \simeq E' \psi.
	\end{align}
	Therefore, the Hamiltonian $\mathscr{H}^*$ is the effective Hamiltonian of $\mathscr{H}_0$ in the non-relativistic limit.
	It should be emphasized that the g-factor of $\mathscr{H}^*$ is written by the cyclotron mass $\mc$, namely, the spin mass of $\mathscr{H}^*$ is exactly the same as its orbital mass.
	This is one of the most important signatures of a Dirac spectrum.
	$\mathscr{H}^*$ is just the Hamiltonian of free electrons, so that its eigenvalue is :
	\begin{align}
		\epsilon (n, s, k_z) &= \left(n + \frac{1}{2}\right) \hbar\wc + \frac{\hbar^2 k_z^2}{2\mh} + s g \mub B_z
		\nonumber\\
		&= \left(n + \frac{1}{2} + s \right)\beta B_z+ \frac{\hbar^2 k_z^2}{2\mh}
\end{align}
	where $\beta =\hbar \wc / B_z$.
	Finally, the energy of $\mathscr{H}_0$ is given by :
	\begin{align}
		E = \pm \sqrt{\D^2 + 2\D \epsilon(n, s, k_z)} ,
\end{align}
	where the ``$+$"  and ``$-$" signs correspond to the conduction and valence band, respectively and the origin is set at the center of the gap at $L$ point.

	The Hamiltonian $\mathscr{H}_0$ is obtained by restricting the number of the bands under consideration to two.
	However,  in addition to these two bands, there are other bands and the existence of these outside bands splits the degenerate Landau levels of the Dirac model.
	By means of the perturbation theory, this effect can be described up to the first order of $B$ by \cite{Baraff1965,Maltz1970,dresselhaus,vecchi}
	\begin{align}
		\mathscr{H}_{\rm p} = g' \mub \bm{s}\cdot \bm{B}.
	\end{align}
	Then, the total Hamiltonian of the extended Dirac model is given by $\mathscr{H} = \mathscr{H}_0 + \mathscr{H}_{\rm p}$, and the total energy is by
	\begin{align}
		E_{n, \pm} &=\sqrt{\D^2+2\D \left( n+\frac{1}{2}\pm\frac{1}{2} \right)\frac{\me}{\mc} \beta_0 B
		+ \frac{\hbar^2 k_z^2}{2m_z}}
		\nonumber\\&
		\pm \frac{g'}{2}\frac{\beta_0}{2}B,
	\label{total}
\end{align}
	where $\beta_0 = |e| \hbar/m_{\rm e}c = 0.1158$ meV/T.
	Note that, for the lowest Landau level ($j=0$), the effect of the outside bands appears in more complex form as is discussed by Vecchi and co-workers analyzing optical conductivity data\cite{vecchi}.
	However,  in contrast to the optical measurements, DC transport does not probe the evolution of the lowest Landau level.
	Therefore, here we express the effect of outside bands effect just by $\mathscr{H}_{\rm p}$ for $j\ge0$ uniformly.
	We also note that $g'$ can be positive and negative as in the theory of Baraff\cite{Baraff1965}, and Maltz and co-workers\cite{Maltz1970}, while it appears as $|g'|$ in the theory of Vecchi and co-workers\cite{vecchi,Vecchi1974}.
	The present experimental results show that the level cross between $0_{\rm e}^+$ and $1_{\rm e}^-$ cannot be interpreted by $|g'|$.
	The obtained Hamiltonian of the extended Dirac model has a quite similar form as the Hamiltonian used by Alicea and Balents\cite{alicea}, though the fitting parameters are different.

	When $j \me \beta_0 B/\mc \D \ll1$, we can employ ``non-relativistic approximation" as
	\begin{align}
		E_{n, \pm}&\simeq \D+ \frac{\hbar^2 k_z^2}{2m_z}+\left( n+\frac{1}{2} \right)\frac{\me}{\mc} \beta_0 B
		\nonumber\\
		&\pm \frac{1}{2}\left(\frac{\me}{\mc} + \frac{g'}{2}\right) \beta_0 B .
	\end{align}
	%
	%
	Then the ratio of the Zeeman splitting to the orbital splitting is estimated as
	%
	%
	%
	%
	\begin{align}
		\frac{E_{\rm Z}}{\hbar \wc}=\frac{E_{n, +}-E_{n,-}}{E_{n+1, \pm}-E_{n, \pm}}
		&\simeq \frac{\left(\me/\mc+g'/2\right)\beta_0 B}{(\me / \mc) \beta_0 B }
		\nonumber\\&
		= 1+\frac{\mc g'}{2\me}.
		\label{Ez/homega}
\end{align}
	The values of $1+\mc g' /2\me$ are listed in Table \ref{table2}.
	It should be noted that Eq. (\ref{Ez/homega}) is obtained on the basis of the non-relativistic approximation. This is a good approximation only when $j\me \beta_0 B /\mc \D \ll1$.
	Therefore, the estimation based on this approximation is only valid for relatively large $\mc$, e. g. when $\bm{B}//$ trigonal.
	For other configurations its domain of validity is restricted to the low magnetic fields.

	If we consider the angular dependence of $\bm{B}$, it is very useful to describe $\mc$ in terms of the mass tensor $\bm{m}$ as
	\begin{align}
		\mc = \sqrt{\frac{\det \bm{m}}{\mh}},
	\end{align}
	in which the longitudinal effective mass $\mh$ is found by taking the component of $\bm{m}$ along the magnetic field,
	\begin{align}
		\mh = \bm{b}\cdot \bm{m} \cdot \bm{b},
	\end{align}	
	where $\bm{b}$ is a unit vector along the magnetic field.
	The effective mass tensor for electrons at $L$ point is given by
	\begin{align}
		\bm{m} = \me
		\begin{pmatrix}
			m_1 & 0 & 0\\
			0 & m_2 & m_4 \\
			0 & m_4 & m_3
		\end{pmatrix},
\end{align}
	where 1, 2, and 3 refer to the binary, bisectrix, and trigonal axes, respectively.
	The tilted angle $\theta$ is written as $\tan 2\theta = 2 m_4 /(m_2 + m_3)$.
	The other two electron pockets are obtained by rotations of $\pm 120^\circ $ about the trigonal axis.
	The additional g-factor is given in the same manner as
	\begin{align}
		g' &= \bm{b}\cdot \bm{g'} \cdot \bm{b}, \\
		\bm{g'} &=
		\begin{pmatrix}
			g'_1 & 0 & 0\\
			0 & g'_2 & g'_4 \\
			0 & g'_4 & g'_3
		\end{pmatrix}.
\end{align}
	
	Now we discuss the energy dispersion of holes at $T$ point.
In a Face-Centered-Cubic (FCC) lattice structure, the two $T$ points are strictly equivalent to the six $L$ points.
	It is natural, therefore, to consider the holes at $T$ point within the same framework as the electrons at $L$ point.
	Note, however, that the band gap at $T$ point is very large $\D_T$ ($\gtrsim 200$ meV\cite{golin,bate,verdun}), which is an order of magnitude larger than the gap at $L$ point. Therefore, the non-relativistic treatment of holes becomes a reasonable approximation.
	Then the hole energy can be expressed like $\mathscr{H}^*$ for the electrons.
	Adding the additional Zeeman term, which originates from the outside bands, we obtain for holes an expression similar to the one used in the SBR model \cite{smith,sharlai}:
	\begin{align}
		E_0 + \D - E =  \left( n+ \frac{1}{2} \right)\hbar \Oc + \frac{\hbar^2 k_z^2}{2\Mh} \pm \frac{G}{2} \mub B,
\end{align}
	where $E_0$ is the hybridization energy.
	The cyclotron frequency ($\Oc$), the longitudinal- ($\Mh$) and cyclotron-mass ($\Mc$), and the g-factor ($G$) for holes is given in the same manner as for electrons.
	The Zeeman splitting energy of this model is:
	\begin{align}
		E_{\rm Z} = G\mub B,
\end{align}
	and so the ratio to the orbital splitting is given by:
	\begin{align}
		\frac{E_{\rm Z}}{\hbar \wc} = \frac{G\Mc}{2\me}.
\end{align}
	This time, this formula gives good approximation for whole direction and amplitude of $\bm{B}$, since the holes are non-relativistic, unlike in the case of electrons, Eq. (\ref{Ez/homega}).

	The effective mass tensor has the form
	\begin{align}
		\bm{M}= \me
		\begin{pmatrix}
			M_1 & 0 & 0 \\
			0 & M_2 & 0 \\
			0 & 0 & M_3
		\end{pmatrix}.
\end{align}
	The effective g-factor is defined in terms of the spin mass $M_{\rm s}$ by
	\begin{align}
		G = 2 \me\sqrt{\frac{\bm{b}\cdot \bm{M}_{\rm s}\cdot \bm{b}}{\det \bm{M}_{\rm s}}}.
	\end{align}
	This effective g-factor already includes the effect of the outside bands, which is closer to the initial two hole bands than that for electrons\cite{luli}.
	So, the effective g-factor of holes may be much larger than that of electrons.

	The Fermi level $\Ef$ is determined by the charge neutrality condition:
	\begin{align}
		\sum_{i=1}^3 N_{e, i}(\Ef) = N_{h}(\Ef),
\end{align}
	where $N_{e, i}$ and $N_h$ are the carrier numbers of electrons at $i$-th pocket and holes, respectively.
	The mass and g-factor tensors used in the present work are listed in Table \ref{electron mass}. The components of the mass tensor are comparable but not identical to those previously reported\cite{smith,edelman,liu}. They have been tuned to give the best agreement with the experimental data obtained in this study.  The parameters $\D = 7.65$ meV and $E_0 = 38.5$ meV are the same used in the SBR model\cite{smith,sharlai}.


\section{The experimental probe and set-up}
In the vicinity of the quantum limit, Nernst effect becomes extremely sensitivity to quantum oscillations\cite{zhu2}. Giant oscillations of the Nernst response have been observed in three semi-metals: bismuth\cite{mangez,behnia2}, Bi$_{0.96}$Sb$_{0.04}$\cite{banerjee} and graphite\cite{zhu1}. Each time a Landau level intersects with the chemical potential, Nernst signal sharply peaks. Recent theoretical works\cite{bergman,sharlai2} have explained the large magnitude of the quantum oscillations in this context.

The giant oscillations dominate a monotonous background which is unusually large. Indeed, the magnitude of the low-field Nernst coefficient in bismuth peaks to the large value of $7 meV K^{-1}$ at about 3 K\cite{behnia1}. This is three orders of magnitude larger than the typical vortex Nernst signal of a type II superconductor. Thus, a Nernst experiment on a bismuth single crystal consists in measuring a relatively large transverse voltage of a few microvolts  generated by a temperature difference as small as few miliKelvins.

We employed a standard one-heater-two thermometers setup to measure the Nernst effect. In our range of investigation, phonon thermal conductivity exceeds the electronic heat conductivity by several orders of magnitude. As a consequence, the change in thermal conductivity induced by magnetic field is negligible and the field dependence of the Nernst coefficient is set by the variation in the transverse electric field as the field is scanned in presence of a constant heat current. The transverse voltage was measured by an EM electronics Nanovoltmeter and collected through copper wires with no junction on their way between room temperature and 4.2K. In this way, thermoelectric noise was suppressed to measure a DC voltage of the order of 1 nV. Measurements were done in a dilution refrigerator inserted in a 12 T superconducting magnet.

The relative orientation of the crystal and the magnetic field was controlled using two coupled Attocube rotators.  A first rotator allowed to explore a window of $\pm60$ degrees in the principal rotational plane. A second smaller rotator allowed to rotate in a plane perpendicular to the principal plane of rotation and was used to adjust the plane of rotation. Angles were determined by a pair of perpendicular Hall probes.  In this study we exclusively focus on the angular evolution of the field position of the Nernst peaks in order to extract the angle-dependent spectrum. We have repeated the experiment on several crystals to check the reproducibility of the results. A Nernst voltage is the vector product of the thermal gradient and the magnetic field. In these  experiments, the heat current was applied along the rotation axis. Therefore, during the rotation, the thermal gradient was perpendicular to the magnetic field,  but not the electric field measured by a set of two fixed electrodes. In order to scan more than 90 degrees, we  measured the electric field along two perpendicular directions using two pairs of electrodes.

\section{Along three high-symmetry axes}
\subsection{Theoretical results}
	
\begin{figure}
\resizebox{!}{0.38\textwidth}{\includegraphics[bb=0 0 733 590]{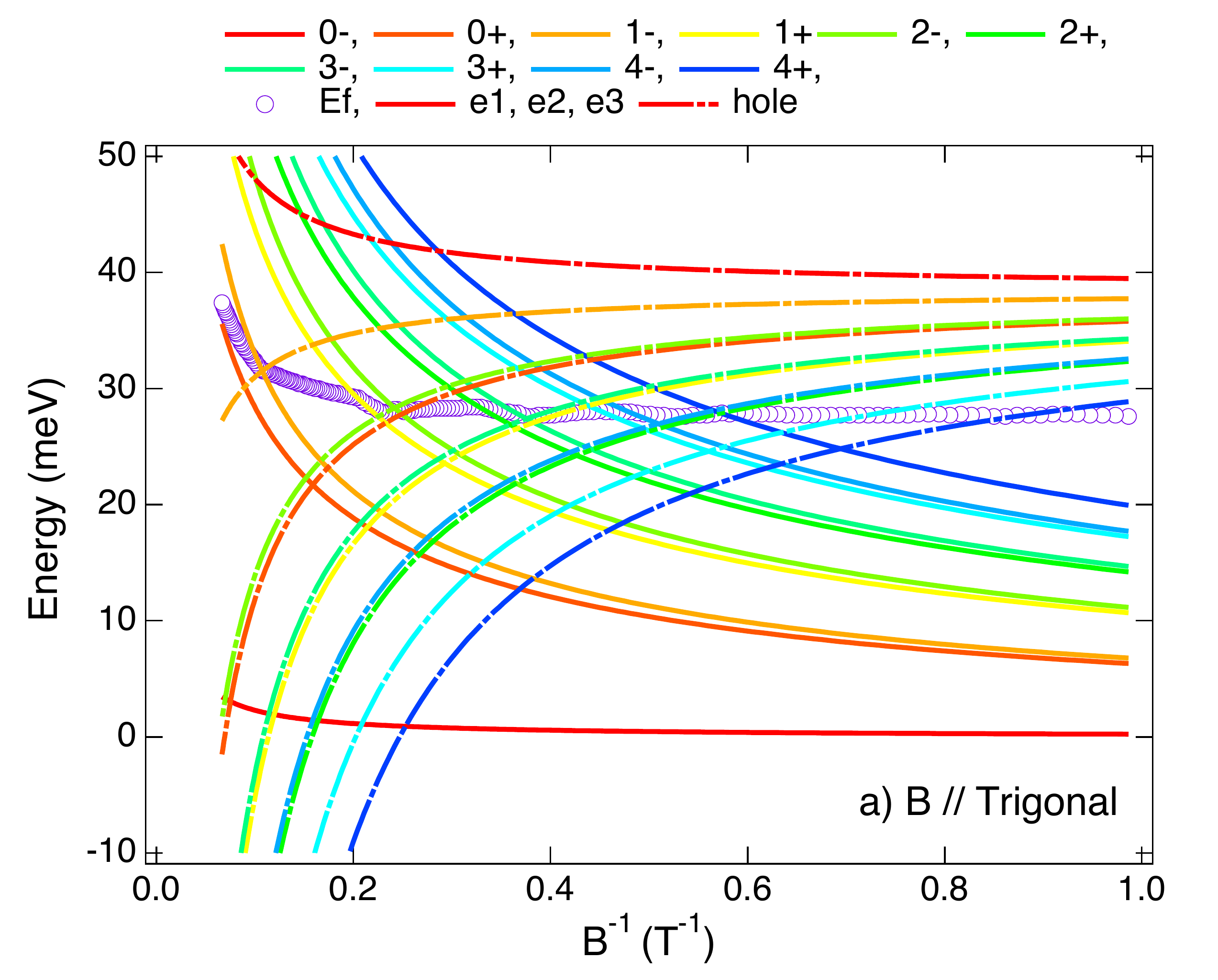}}
\resizebox{!}{0.38\textwidth}{\includegraphics[bb=0 0 733 585]{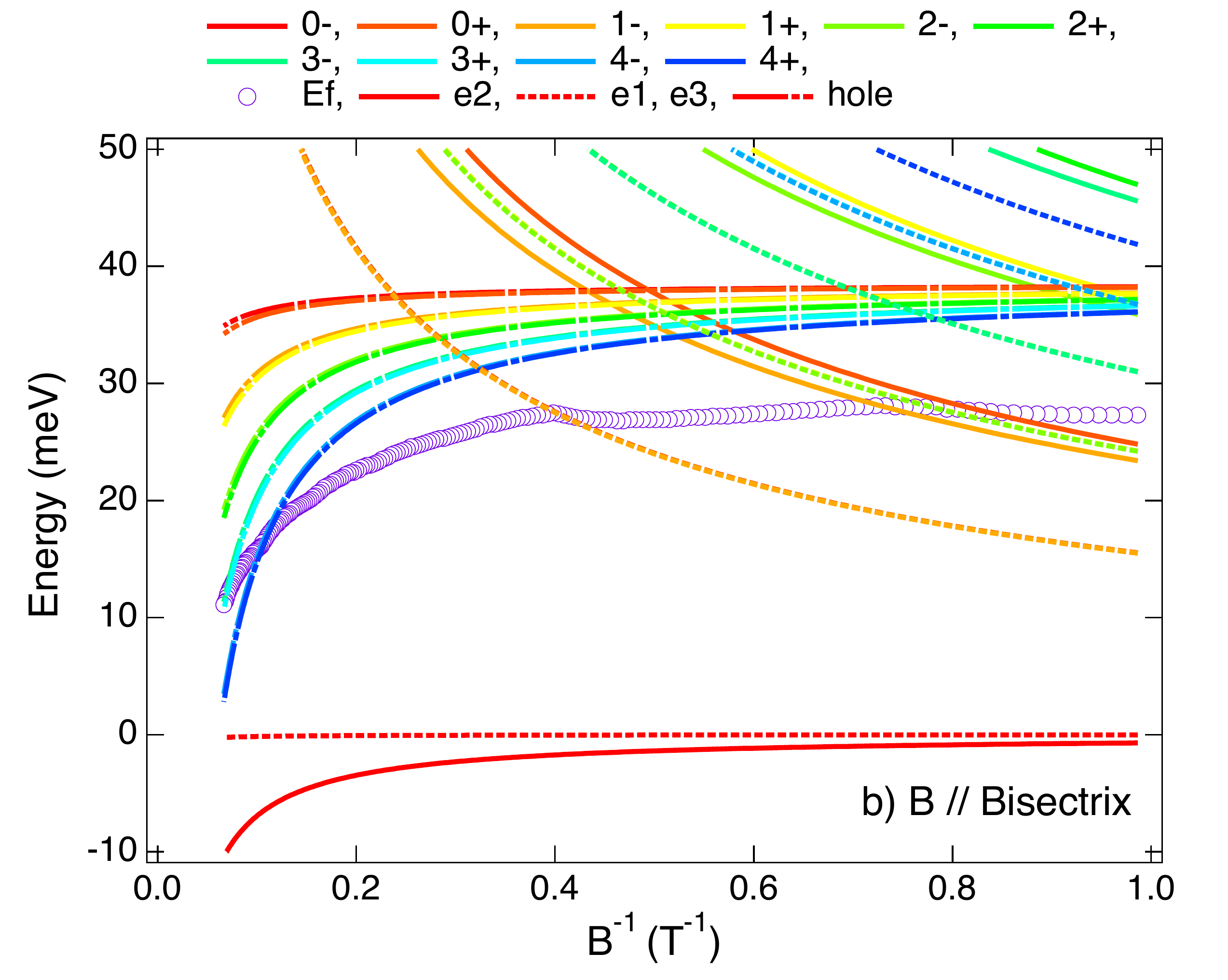}}
\resizebox{!}{0.38\textwidth}{\includegraphics[bb=0 0 733 588]{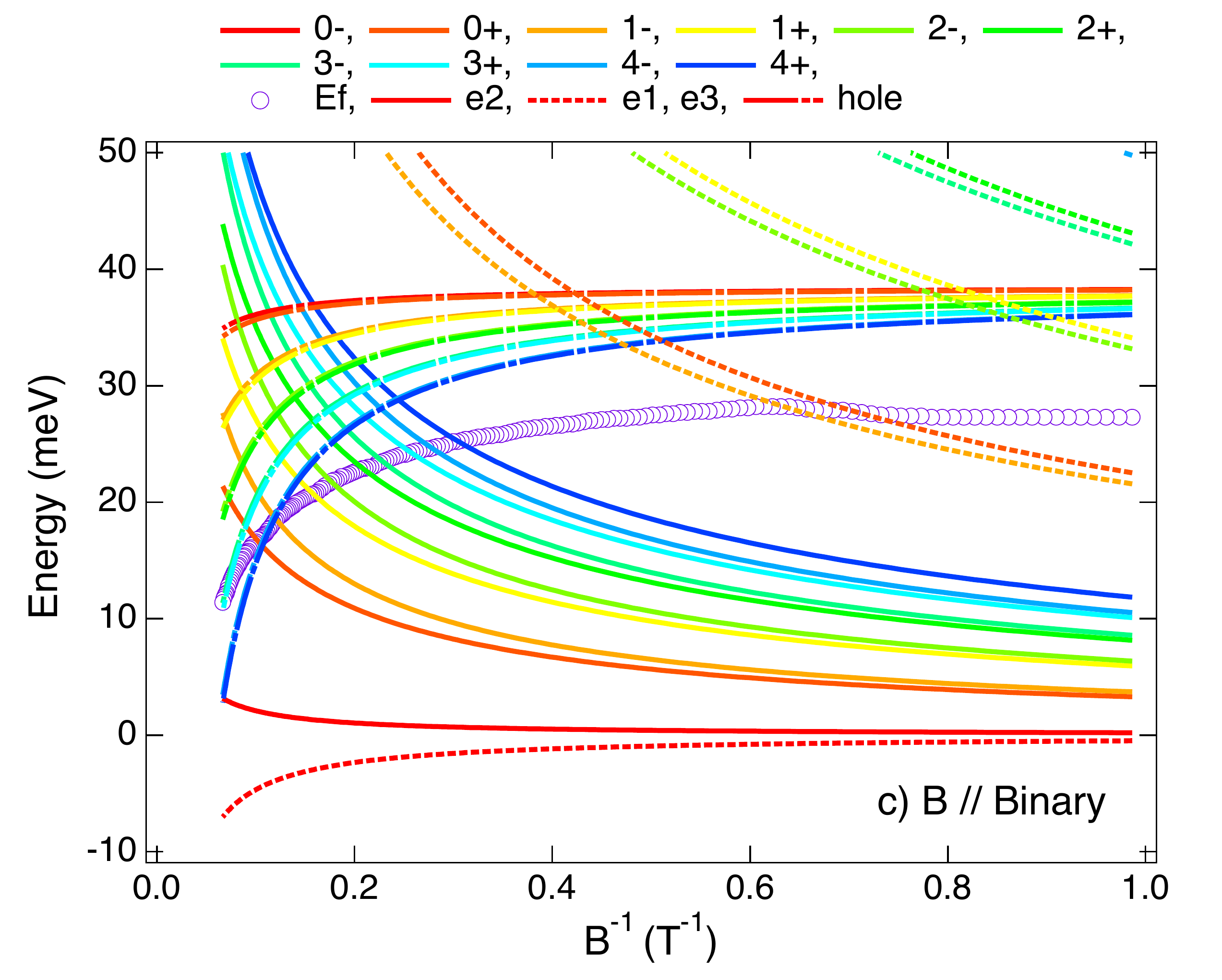}}
\caption{Landau levels ($k_z = 0$) as a function of $B^{-1}$ for a magnetic field along a) Trigonal, b) Bisectrix and c) Binary direction. The origin of the energy is taken at the bottom of the conduction band at zero field.}
\label{E-B^-1}
\end{figure}

The bottom positions of Landau levels as a function of $B^{-1}$ are plotted in Fig. \ref{E-B^-1}. As the field is swept, the levels of electrons move upward and those of holes move downward. The Fermi level shifts with this change in the density of electrons and holes in order to satisfy the charge neutrality.

As seen in different panels of Fig. \ref{E-B^-1}, for $B\parallel$ bisectrix and $B\parallel$ binary, the Fermi level moves downward with increasing $B$, whereas it moves upward for $B\parallel$ trigonal  as a consequence of the rising $0_{\rm h}^-$ level.
\begin{figure}
\resizebox{!}{0.35\textwidth}{\includegraphics[bb=0 0 710 513]{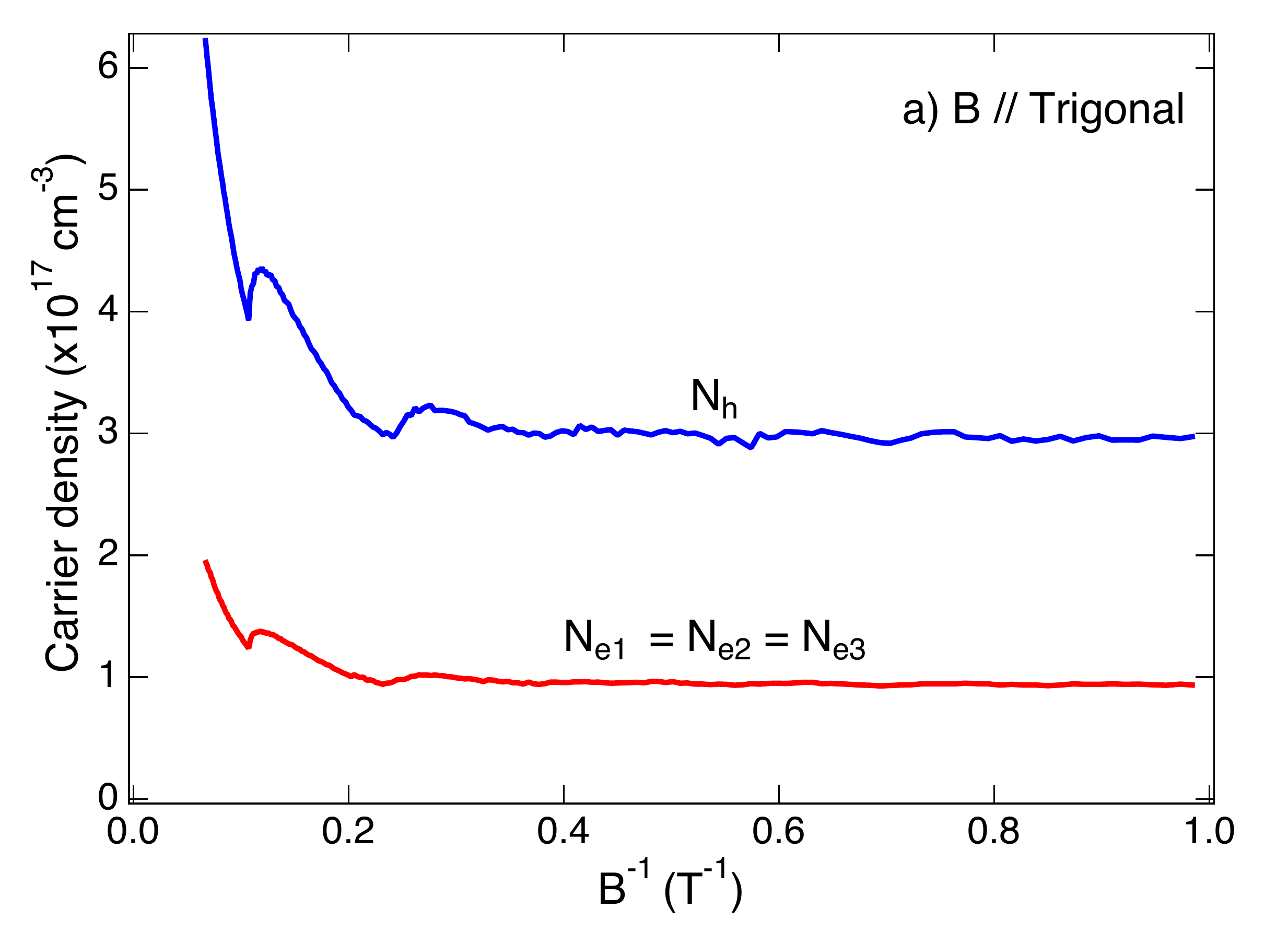}}
\resizebox{!}{0.35\textwidth}{\includegraphics[bb=0 0 710 513]{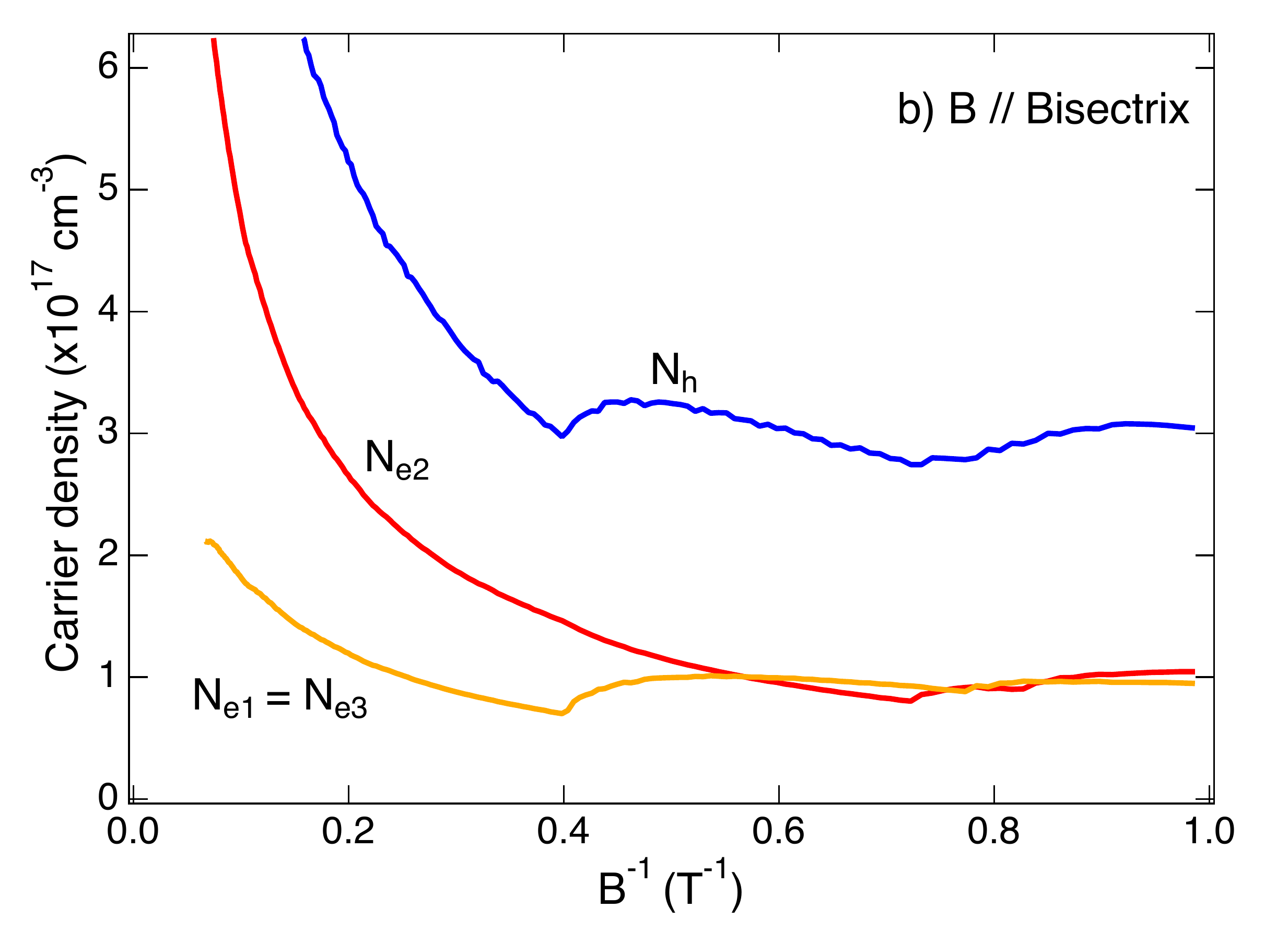}}
\resizebox{!}{0.35\textwidth}{\includegraphics[bb=0 0 710 513]{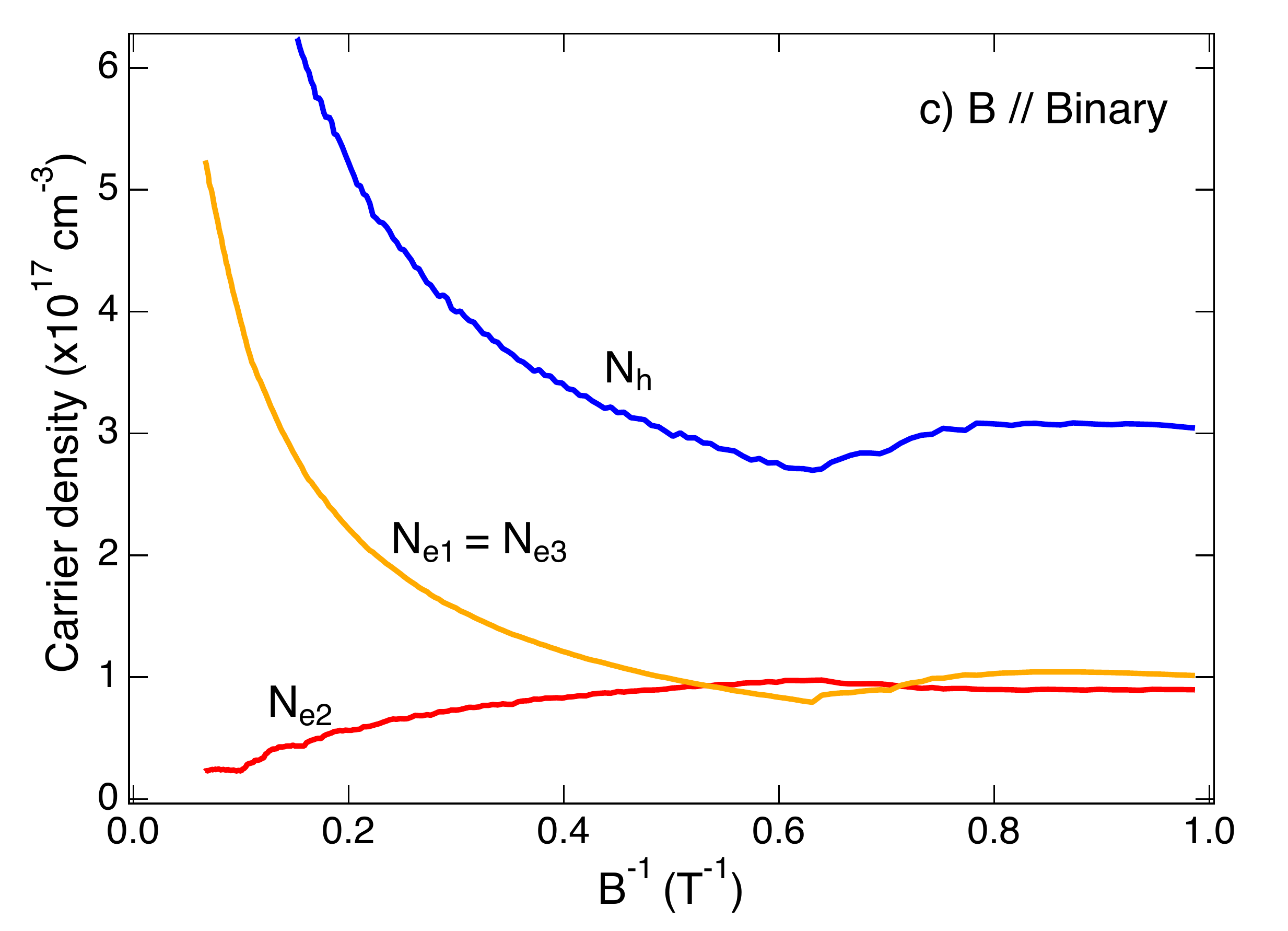}}
\caption{Carrier density as a function of $B^{-1}$ for a magnetic field along a) Trigonal, b) Bisectrix and c) Binary direction. }
\label{N-B^-1}
\end{figure}
	
The variation of carrier density with magnetic field is shown in Fig. \ref{N-B^-1}. The density of carriers basically increases with increasing magnetic field due to the Landau degeneracy. Only for $B\parallel$ binary, the carrier density of the electron pocket along the bisectrix axis decreases with increasing magnetic field  since the level of $0_{\rm e2}^-$ approaches to $\Ef$ as seen in Fig. \ref{E-B^-1}.c).

\subsection{Experimental results}
	
\begin{figure}
\resizebox{!}{0.34\textwidth}{\includegraphics[bb=0 0 725 522]{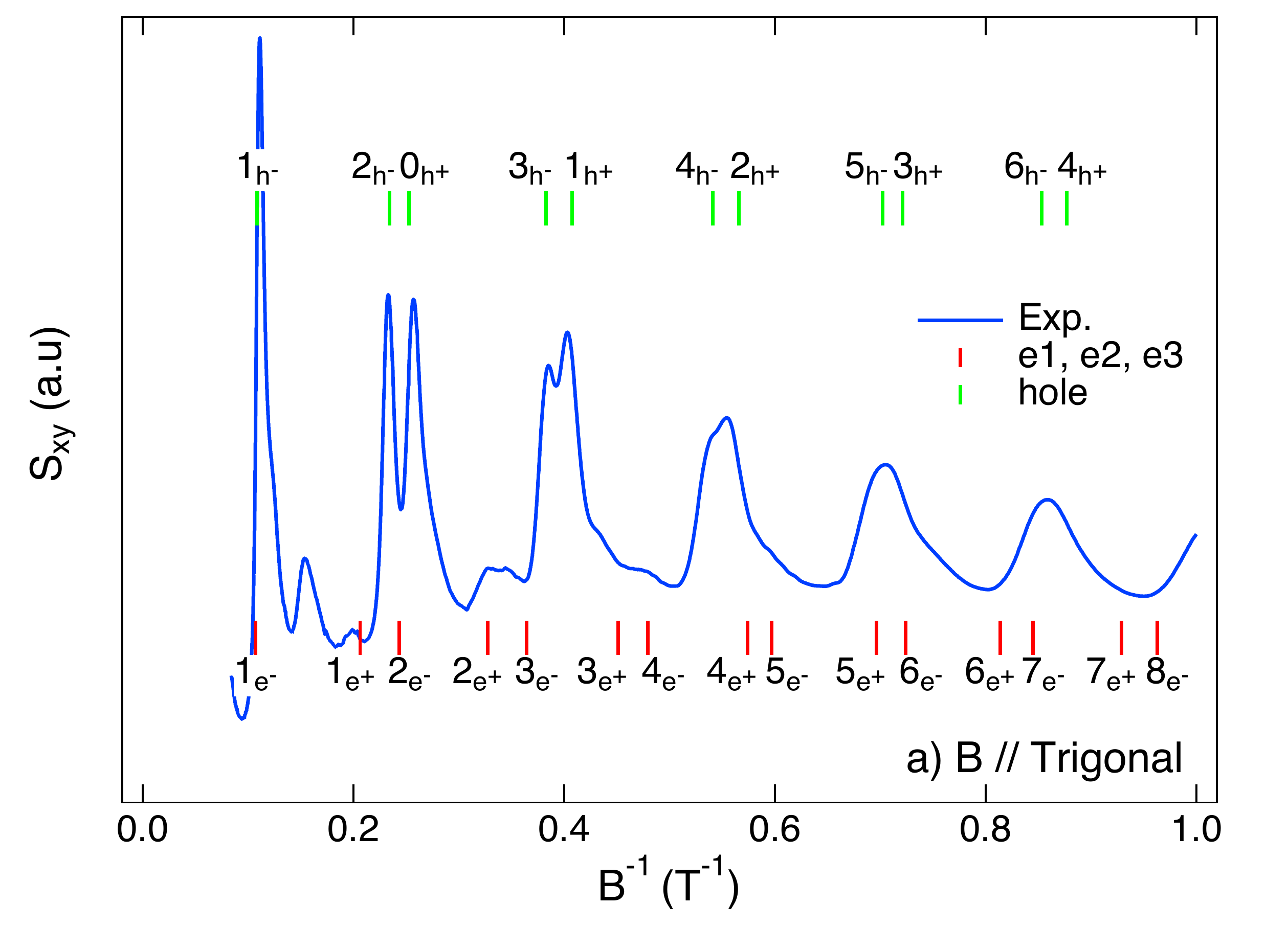}}
\resizebox{!}{0.34\textwidth}{\includegraphics[bb=0 0 733 522]{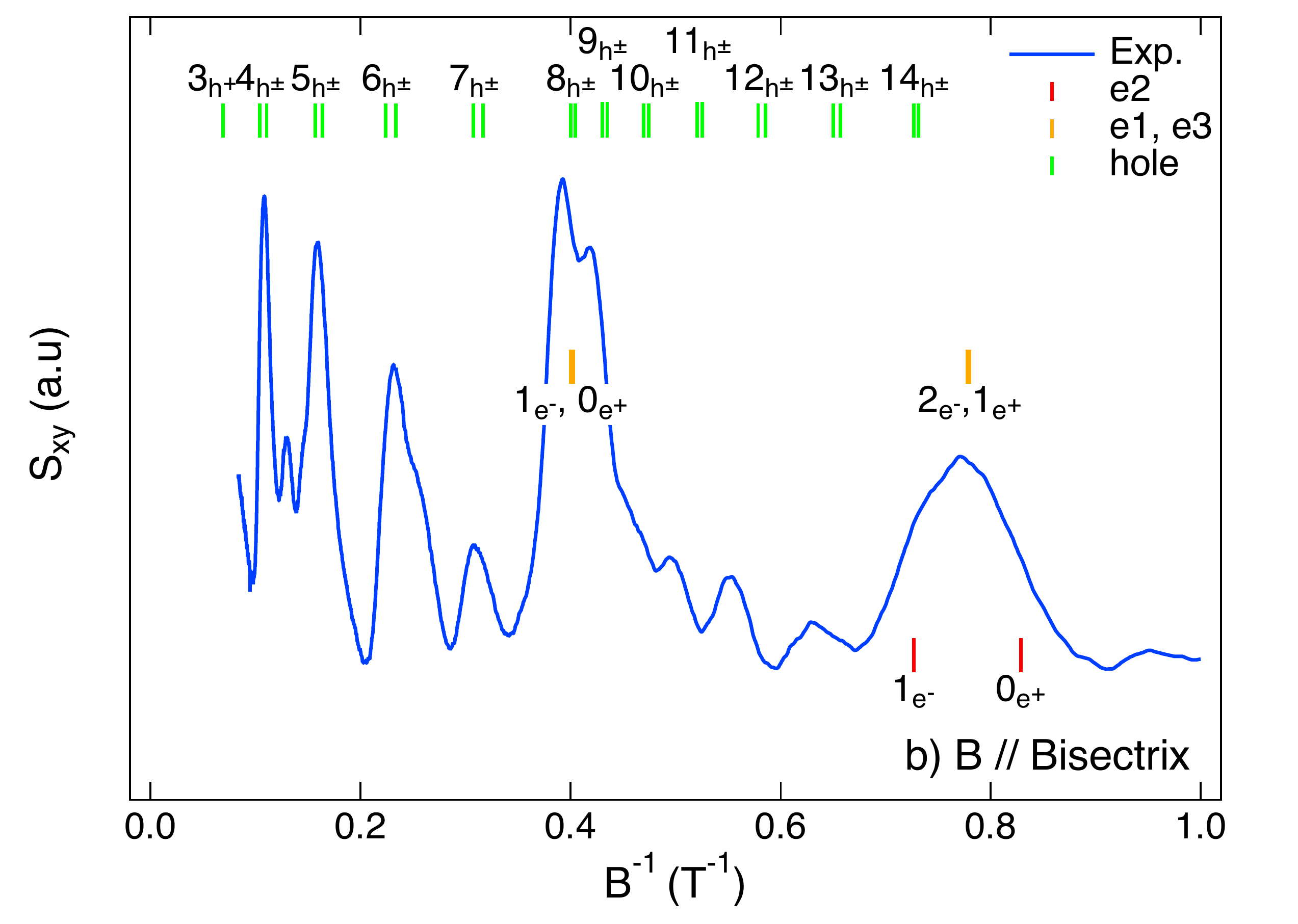}}
\resizebox{!}{0.34\textwidth}{\includegraphics[bb=0 0 725 522]{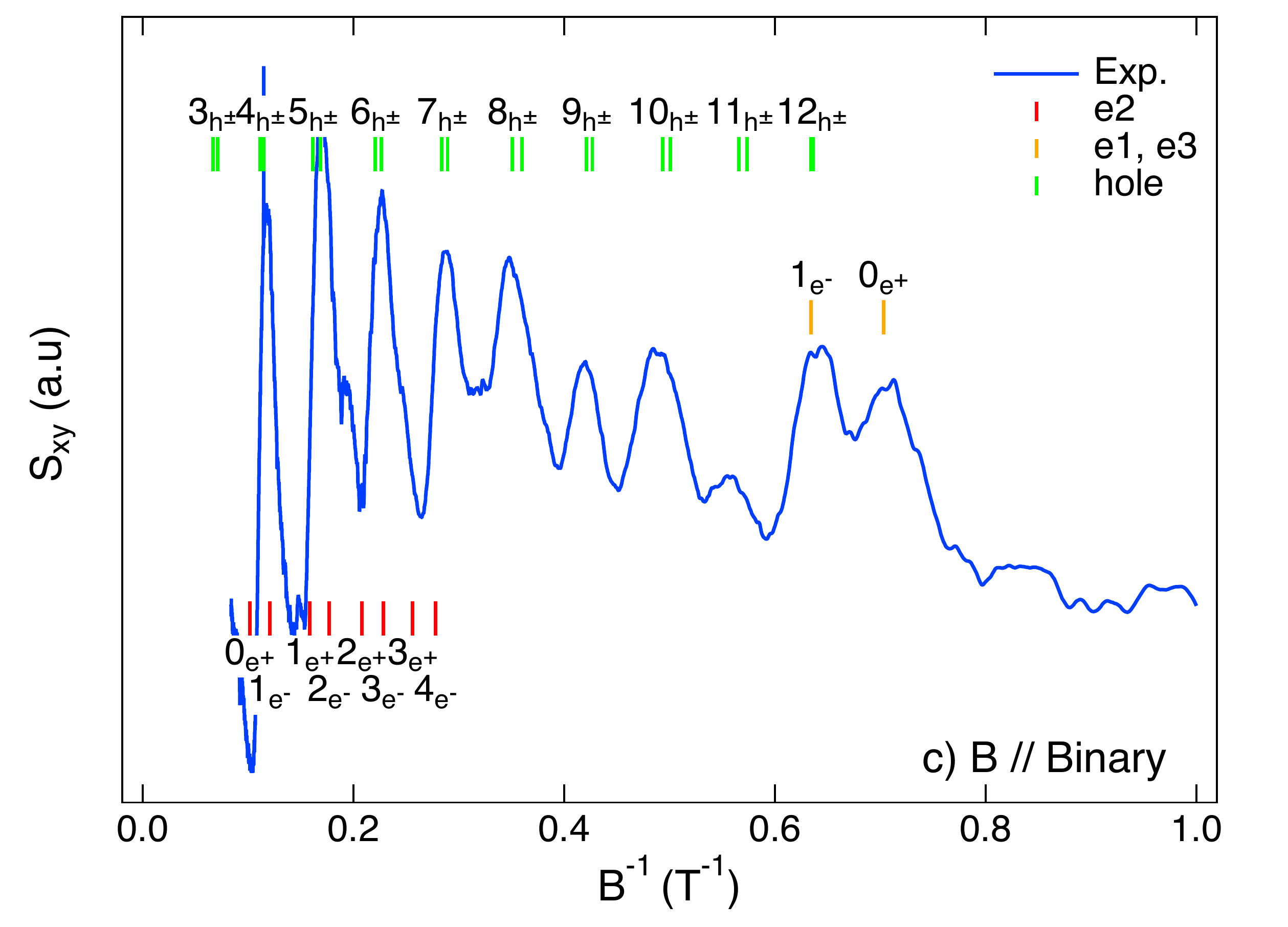}}
\caption{Comparison between theory (vertical lines) and experiment (solid curves) for a magnetic field along a) Trigonal, b) Bisectrix and c) Binary direction. For each configuration, the Nernst voltage is plotted as a function of B$^{-1}$).}
\label{spec}
\end{figure}

In Fig. \ref{spec}, the quantum oscillations of the Nernst response are plotted against the inverse of magnetic field for three different axes. In the same figure the expected theoretical positions for the intersection of the Fermi level with each Landau level is marked by vertical lines.

When the field is along the trigonal axis (the top panel), experiment detects oscillations with a relatively simple pattern as previously reported\cite{behnia2}. The Nernst response is dominated by holes, which give rise to peaks with a periodicity of 0.147 T$^{-1}$. At high enough field ($B>1.5 T$) (that is for low Landau-level numbers), electron peaks become detectable and the Zeeman splitting of holes becomes apparent. The quantum limit of holes is marked by a sharp peak at 9 T (B$^{-1}$=0.11 T). Above this field holes reside in their lowest spin-polarized Landau level and electrons are at their two lowest Zeeman-split Landau level.

When the field is perpendicular to the trigonal axis, the Nernst response is no more dominated by holes.  As seen in the middle panel Fig. \ref{spec}, large Nernst peaks due to electron pockets become visible when the field is along the bisectrix. In this configuration, a field as small as 2.5 T puts the carriers of all three electron pockets in their lowest spin-polarized Landau level (that is $0_{\rm e}^-$). Only above this field, the hole peaks become prominent. When the field is along the binary axis (the bottom panel of Fig. \ref{spec}), the quantum limit of two out of the three electron pockets is attained at 1.5 T. Hole quantum oscillations dominate the spectrum, as the field exceeds this magnitude. The quantum limit of the third electron pocket is attained in a field of about 11 T.

A careful examination of Fig. \ref{spec} indicates that for both holes and electrons, the agreement between experiment and the theory is rather good.

\begin{figure}
\resizebox{!}{0.45\textwidth}{\includegraphics[bb=0 0 472 517]{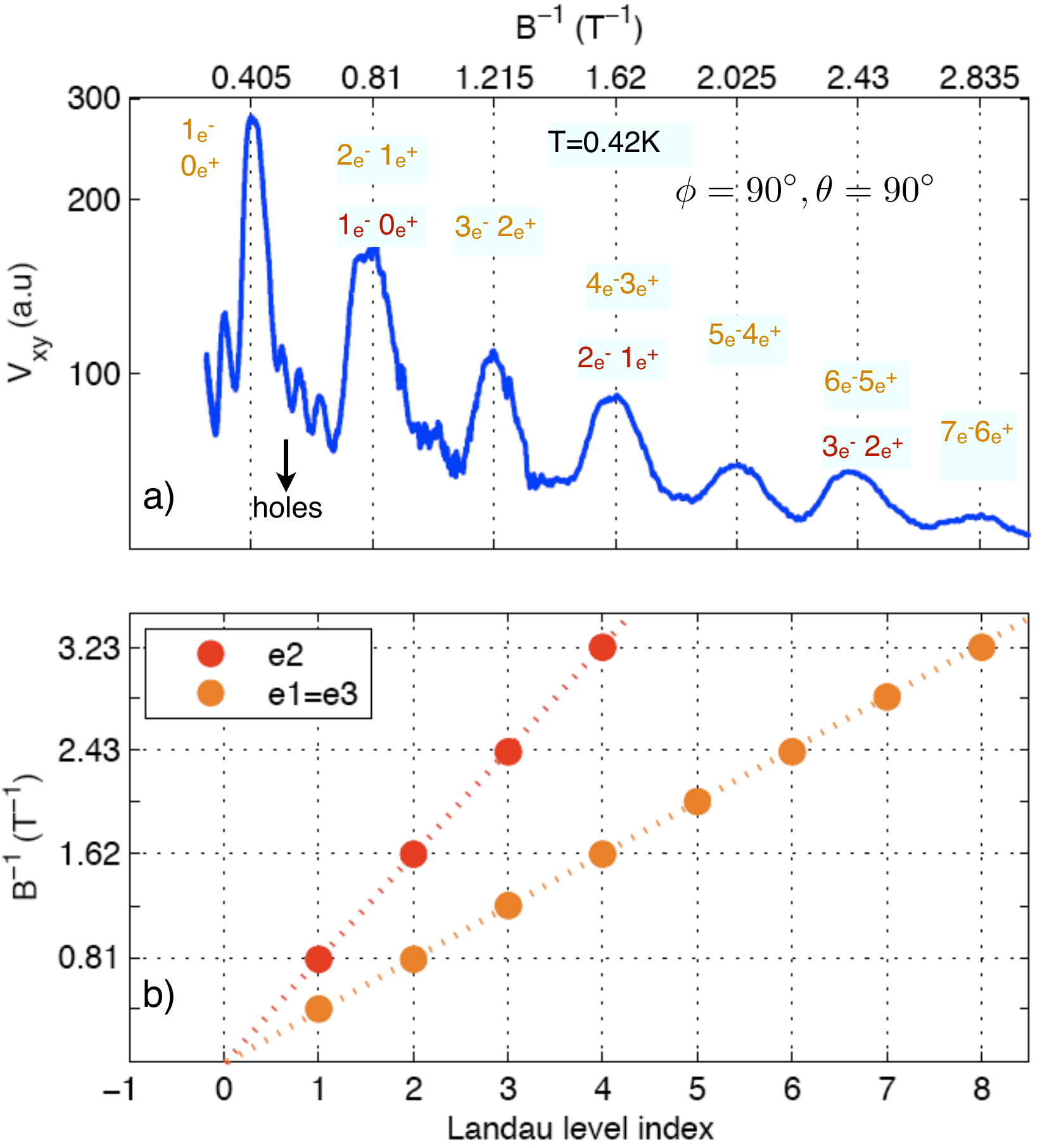}}\caption{Top: Nernst voltage as a function of B$^{-1}$ with the magnetic field along the bisectrix. Electron peaks occur with a periodicity of 0.405 $T^{-1}$. This period is twice the cross section of an electron ellipsoid along the bisectrix axis. When, B$^{-1}$ is an even (odd) number of times this value, a Nernst peak is sixfold (fourfold) degenerate. Bottom, the B$^{-1}$ position of Nernst peaks as a function of their Landau index for the three pockets. Dirac spectrum leads to a vanishing intercept.}
\label{spec_bis_zoom}
\end{figure}

The B $\|$ bisectrix configuration deserves special attention. The upper panel of Fig. \ref{spec_bis_zoom} displays the Nernst voltage as a function of B$^{-1}$ for this configuration in a limited field range ($0.3 T < B < 3 T$) in order to highlight the Dirac spectrum of the electrons. As seen in the figure,  quantum oscillations of the electron pockets are strictly periodic as a consequence of the stability of the chemical potential in this field window. Nernst peaks appear with remarkable regularity at multiples of 0.405 $\pm$ 0.005 T$^{-1}$. The simplicity of this spectrum is a result of two independent factors. First, in the B $\|$ bisectrix configuration, the cross-section of the two pockets, e1 and e3 is exactly twice (that is 1/$\cos(\pi/3)$) the cross section of the pocket e2. Second, in this configuration, the Zeeman energy and the cyclotron energy are almost identical, in other words, the n,+ and n+1,- states become almost degenerate.

As a consequence of these two features, Nernst peaks for e2 emerge at multiples of 0.81 T$^{-1}$ and those for the pockets e1 and e3 at multiples of 0.405 T$^{-1}$. Therefore, the experimentally resolved peaks at odd or even multiples of 0.405T$^{-1}$ are respectively fourfold and sixfold degenerate.

As seen in the Table II, when B $\|$ bisectrix, the theoretical magnitude of 1+$\frac{\mc g'}{2}$ remains close to unity leading to a spectrum very close to a purely Dirac spectrum in agreement with the experiment. As seen in the lower panel of Fig. \ref{spec_bis_zoom}, when the B$^{-1}$ position of the Landau levels are plotted \emph{vs.} the index number, the intercept is very close to zero as expected in the case of a the Dirac spectrum.

We note that in this configuration, when the field exceeds 2.5 T, electron-like carriers will reside in their lowest spin-polarized Landau level.

\section{Angle-resolved Landau spectrum}

\begin{figure}
\resizebox{!}{0.45\textwidth}{\includegraphics[bb=0 0 524 502]{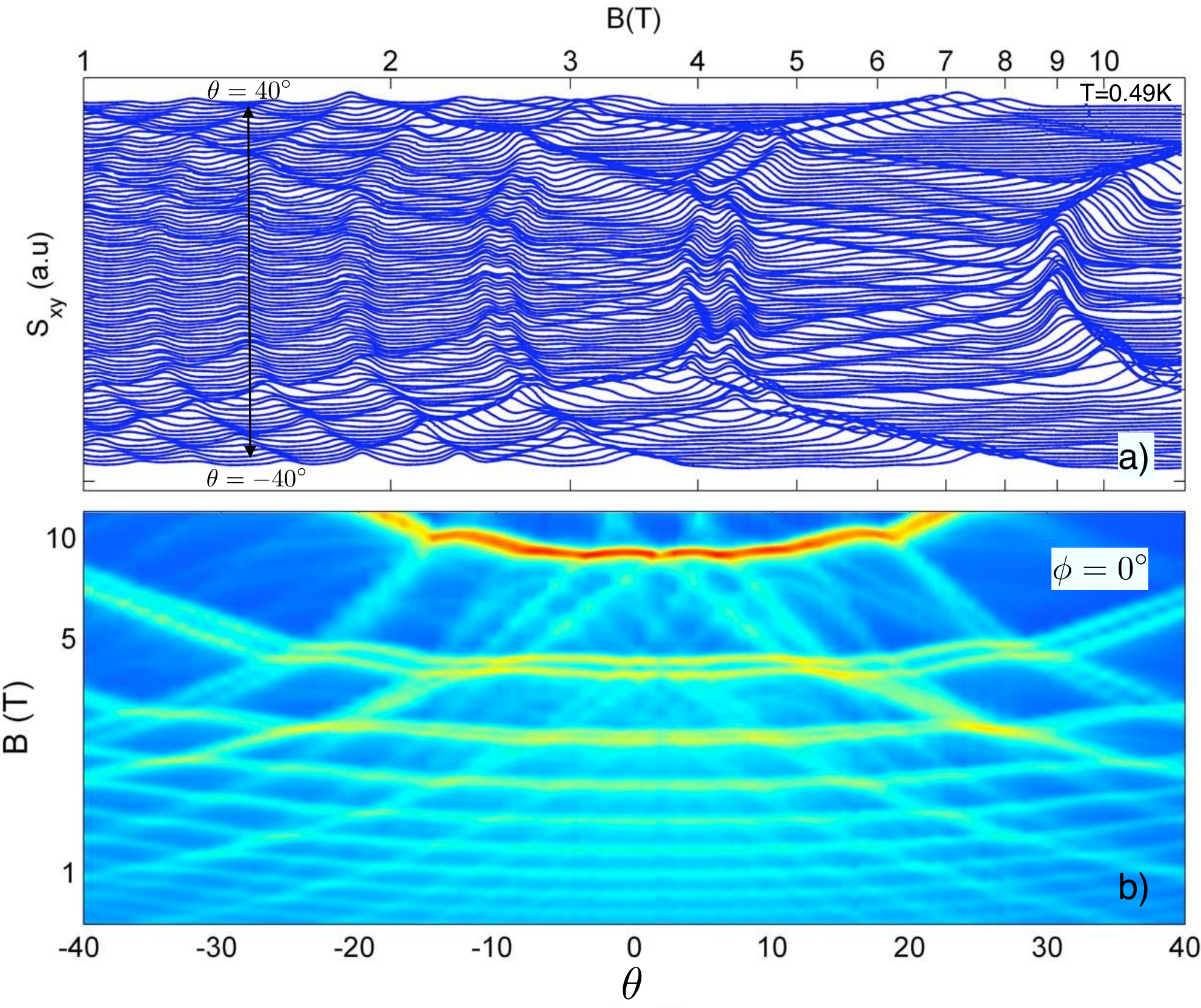}}\caption{a) The evolution of the field dependence of the Nernst signal S$_{xy}$ with the orientation of the magnetic field as the field rotates in the (trigonal, binary) plane. Curves are shifted for clarity.
b) Color map of the data presented in the upper panel. Bright lines track the angular evolution of the Nernst peaks corresponding to the intersection of a Landau level and the chemical potential.}
\label{map}
\end{figure}

The angle-dependent Nernst effect was studied for a magnetic field rotating in  three perpendicular planes of rotation. Fig. \ref{map} presents a typical set of Nernst data. The upper panel (Fig. \ref{map}.a) presents the Nernst signal, S$_{xy}$ as a function of magnetic field for different field orientations tilted off the trigonal axis from -40 degrees to 40 degrees. Quantum oscillations are clearly visible. Each peak corresponds to the intersection of a Landau level of one of the pockets with the chemical potential. The position of peaks shifts as the magnetic field rotates. A color plot of the same data is presented in the lower panel and reveals the Landau spectrum. The angular evolution of the Nernst peaks plots bright lines which take the shape of a network.  The quasi-horizontal lines which curve upward as the field is tilted off the trigonal axis correspond to the Landau levels of the hole pocket: as the field is tilted, the cross section of this ellipsoid increases. The quasi-perpendicular lines correspond to the three electron pockets, which lay almost perpendicular to the hole pocket.

\begin{figure}
\resizebox{!}{0.45\textwidth}{\includegraphics[bb=0 0 554 503]{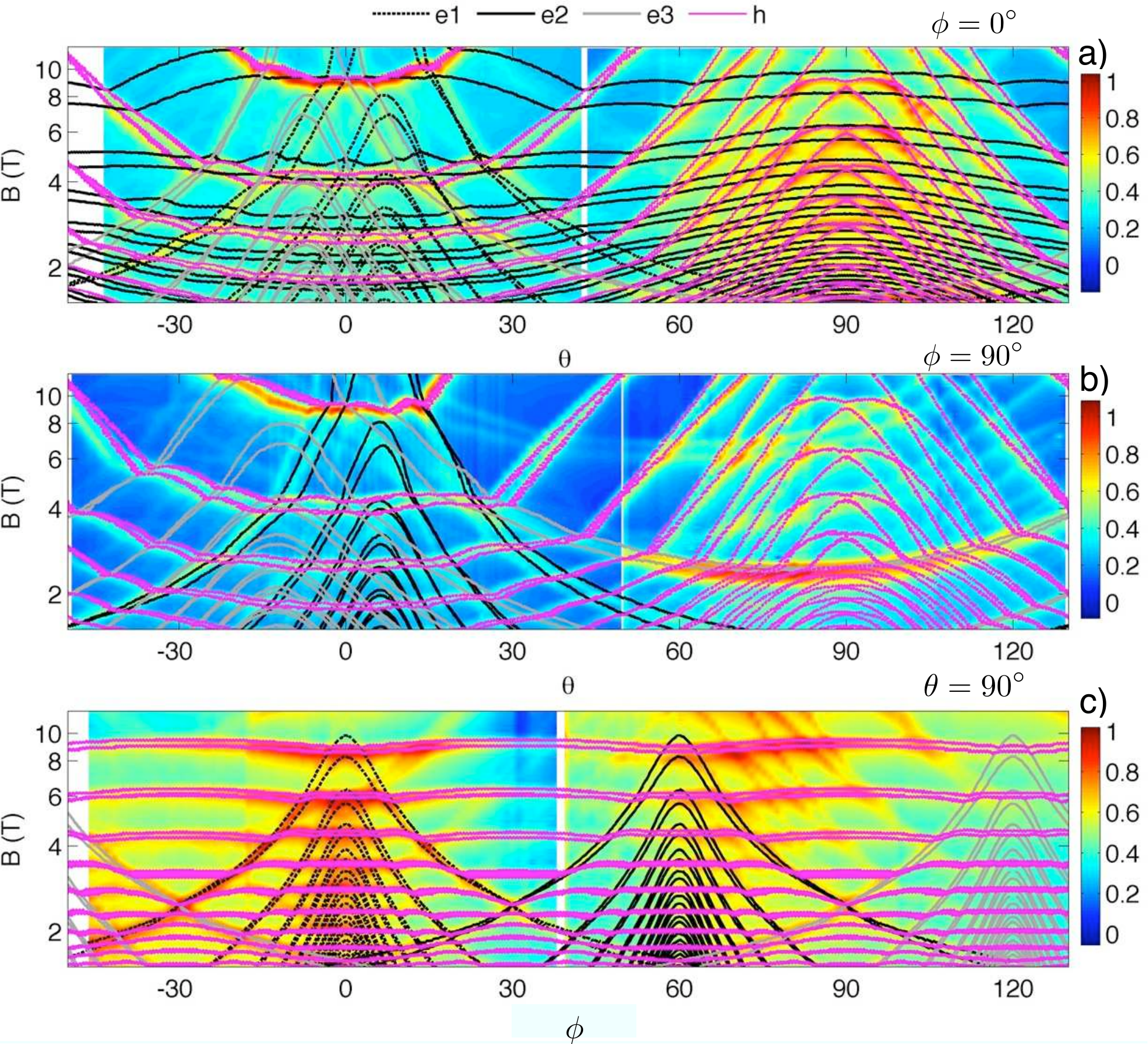}}
\caption{  Color map of S$_{xy}$ when the magnetic field rotates in three perpendicular planes:  a) (Trigonal,Binary) plane (i.e. $\phi=0^\circ$), b) (Trigonal, Bisectrix) plane (i.e. $\phi=90^\circ$) and c) (Binary, Bisectrix) plane (i.e. $\theta=90^\circ$). On top of the color map, theoretical lines corresponding to the hole (purple) and the three electron pockets, e$_1$ (dotted), e$_2$ (black solid) and $e_3$ (gray solid) are plotted. In each panel there are two parts corresponding to the data obtained with two different pair of electrodes. }
\label{fig:AllPlanes}
\end{figure}

 We used the same procedure to determine the color map of the Nernst signal in three principal planes. The results are presented in Fig. \ref{fig:AllPlanes}. In each panel, there are two adjacent parts representing the data obtained from different pair of electrodes on the same sample.  No  discontinuity appears as one crosses from one set of electrodes to the other. Theoretical lines derived from the model discussed above are put on the top of the color plots.  As seen in the figure, these lines are in rough agreement with the experimental data.

\begin{figure}
\resizebox{!}{0.45\textwidth}{\includegraphics[bb=0 0 540 550]{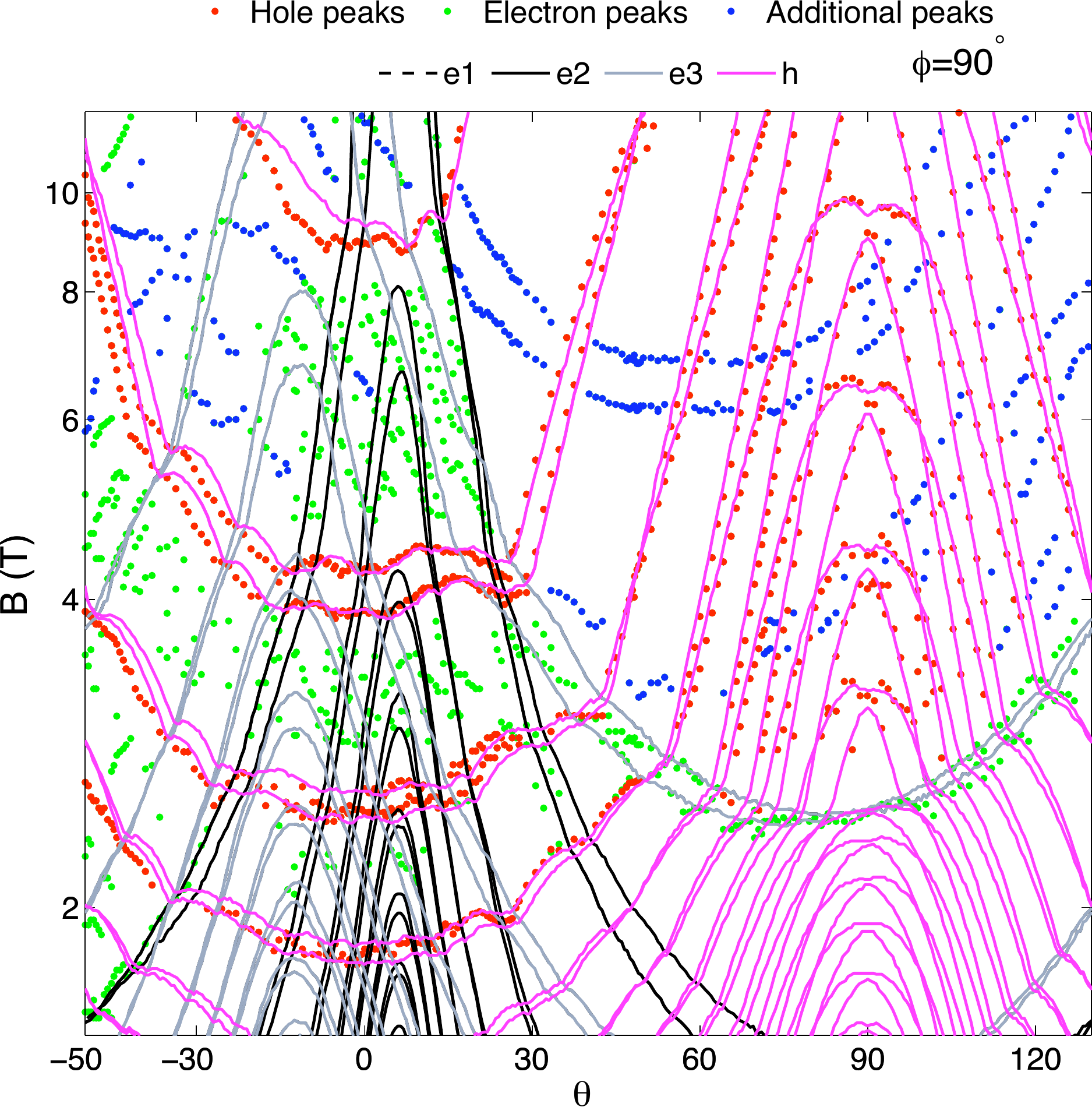}}\caption{The Landau spectrum for a magnetic field rotating in the (trigonal, bisectrix) plane  according to theory (lines) and experiment (symbols). Additional peaks refer to those which can be clearly identified as not associated with an electron Landau sub-level.}
\label{trigbis}
\end{figure}

The three following figures allow a comparison between the theory and experiment for each of the three rotating planes. In each figure, the angular variation of the field at which a Landau band crosses the chemical potential is compared to the angular variation of the fields at which the Nernst signal peaks.

It is easy to distinguish the Nernst peaks associated with holes, since there is a single hole pocket and their intensity evolves continuously with rotation. As seen in the figure, the theoretical model gives a very good account of the experimentally-resolved hole peaks in all of the three planes.
When the field rotates in the (binary, bisectrix) plane (Fig. \ref{binbis}), the hole mass remains constant with no angular dependence (See Table  II). However, as a consequence of a finite angular variation of the Fermi energy, which is pulled by the change in the density of electron-like carriers, the hole lines present a small angular modulation. As seen in the figure, this theoretically expected angular variation is resolved by the experiment. Note also the doubling of the experimentally resolved hole peaks, which is a consequence of a slight misalignment.

\begin{figure}
\resizebox{!}{0.45\textwidth}{\includegraphics[bb=0 0 540 550]{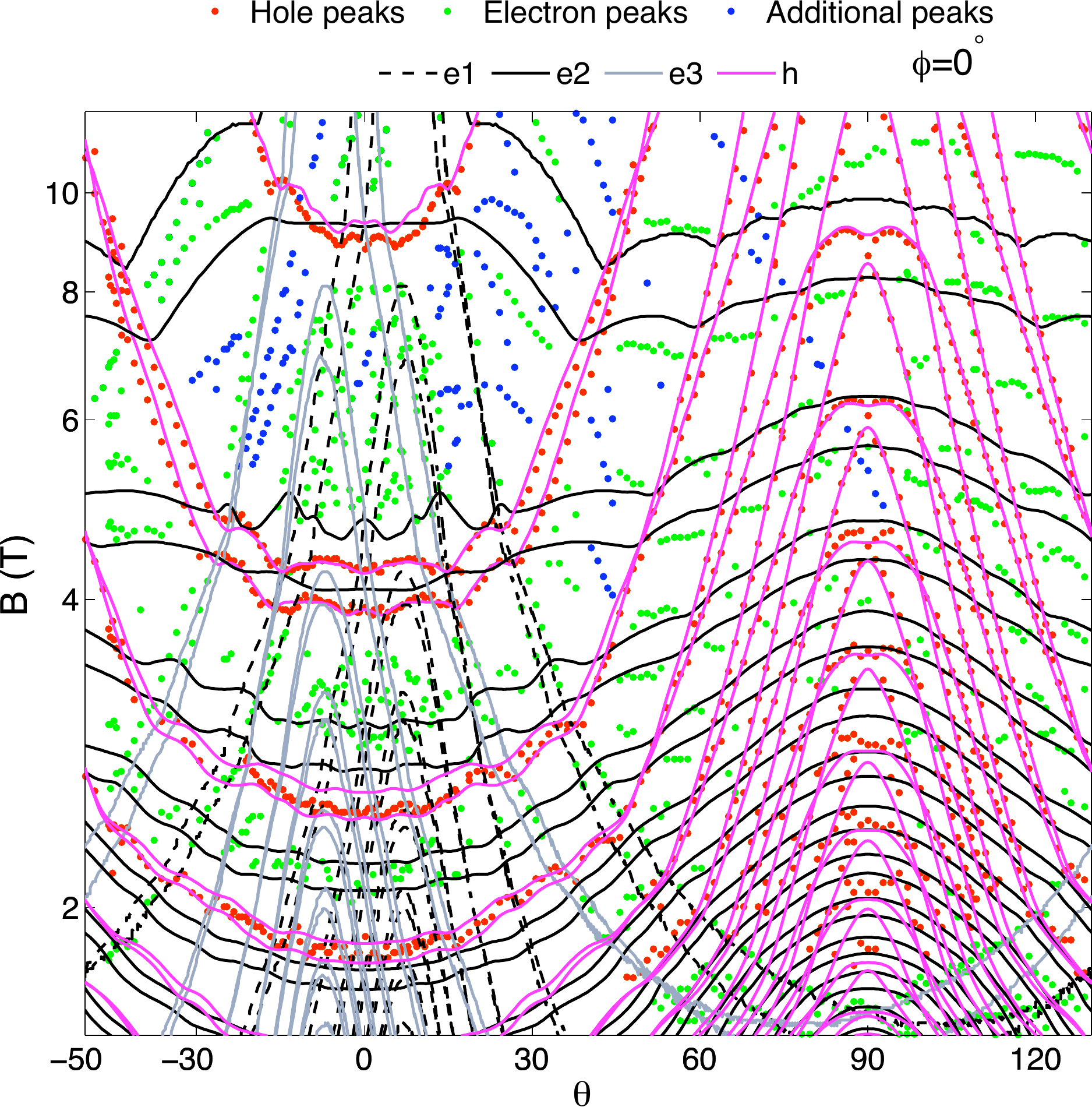}}\caption{ The Landau spectrum for a magnetic field rotating in the (trigonal, binary) plane  according to theory (lines) and experiment (symbols). Additional peaks refer to those which can be clearly identified as not associated with an electron Landau sub-level.}
\label{trigbin}
\end{figure}

\begin{figure}
\resizebox{!}{0.45\textwidth}{\includegraphics[bb=0 0 540 550]{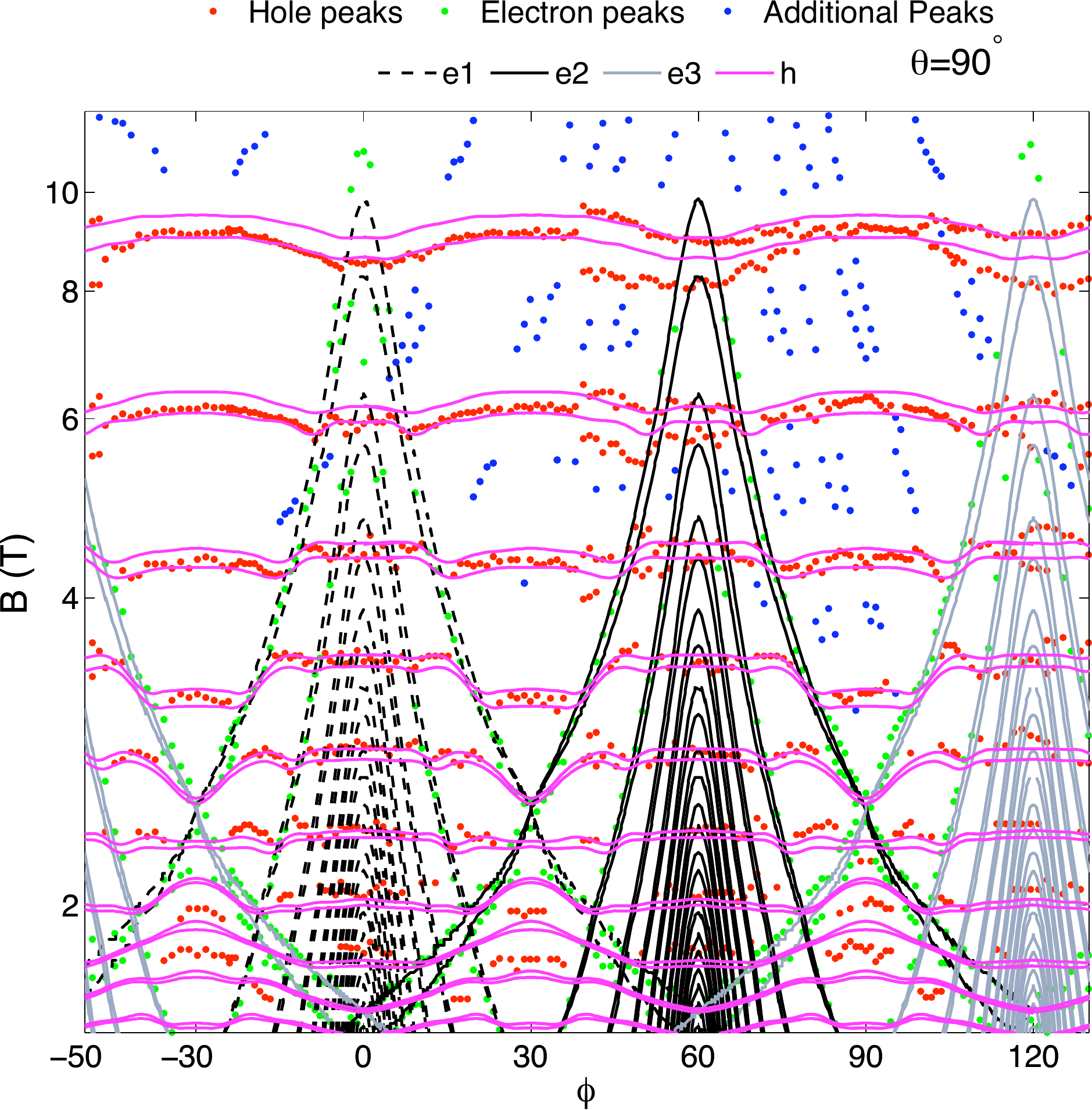}}\caption{ The Landau spectrum for a magnetic field rotating in the (binary, bisectrix) plane  according to theory (lines) and experiment (symbols). Additional peaks refer to those which can be clearly identified as not associated with an electron Landau sub-level.}
\label{binbis}
\end{figure}

As far as the electron lines are concerned, the agreement between theory and experiment is less satisfactory. Note that the three distinct electron pockets become degenerate only when the field is along the trigonal axis. We found that the angular evolution of those Nernst peaks which can be attributed to electrons is in rough agreement with theory. However, this comparison is complicated by the persistence of additional Nernst peaks of unknown origin\cite{behnia3,yang2}, which are hard to distinguish from the electron peaks. In the three Figures \ref{trigbis},\ref{trigbin},\ref{binbis}, a specific symbol marks those additional peaks. This symbol is used only when the resolved peak is clearly not one of the expected electron peaks. However, at this stage, in this field range and in particular when the field is almost oriented along the trigonal axis, we cannot definitely distinguish between Nernst peaks associated with the electron pockets and those which are unexpected. As the field increases and the number of expected electron lines decreases, it becomes more straightforward to clearly distinguish between theoretically expected phase diagram and additional lines resolved by experiment. This was the case in a recent angle-dependent Nernst study extended to high-magnetic fields by our group\cite{yang2}.

\begin{figure}
\resizebox{!}{0.5\textwidth}{\includegraphics[bb=0 0 477 521]{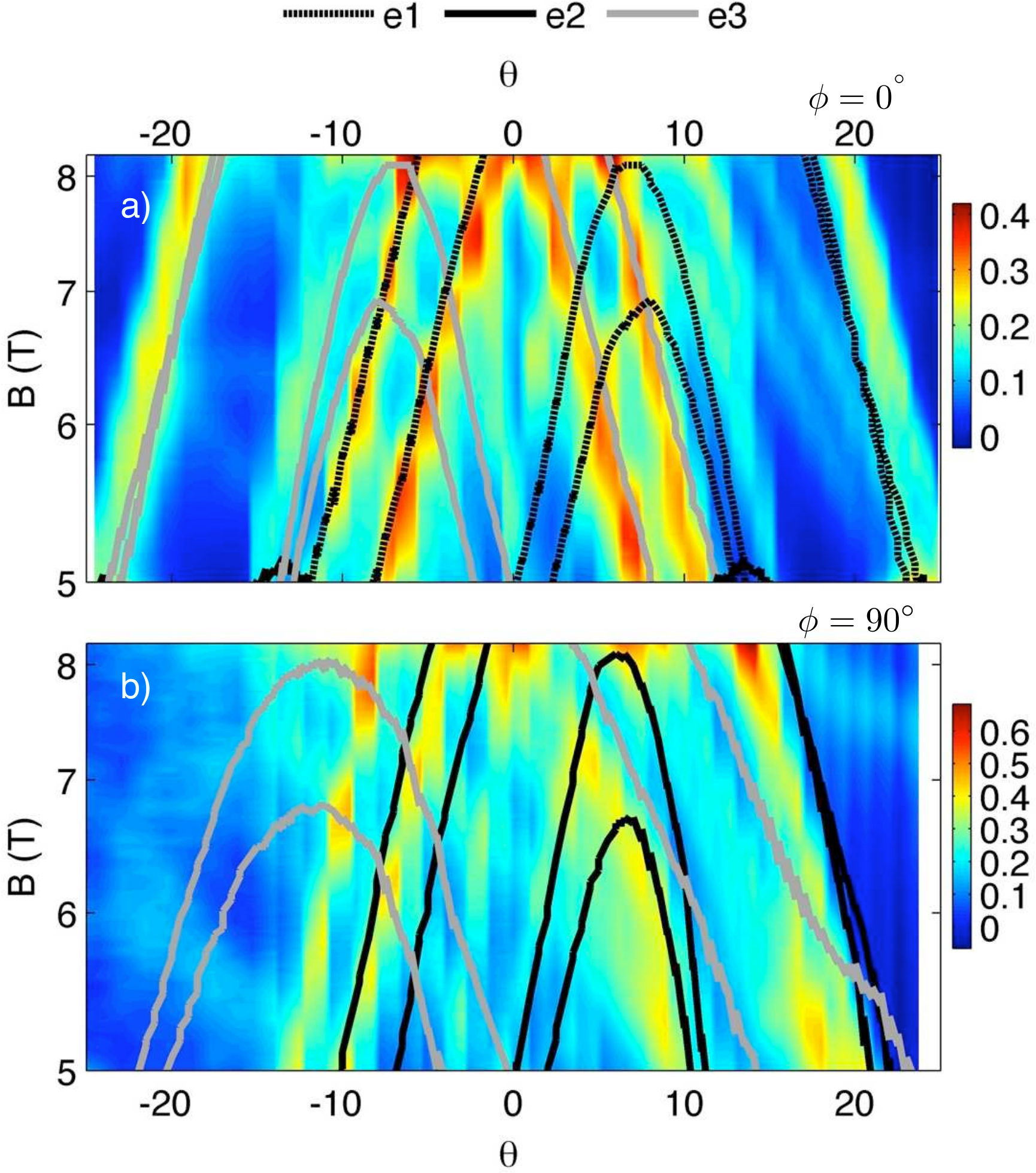}}\caption{ Zoom on Color map of S$_{xy}$ for a magnetic field near the trigonal axis. The magnetic field rotates either  in the (trigonal, binary)  plane ($\phi=0$) (a) or in the plan (trigonal, bisectrix) plane ($\phi=90^\circ$). On top of the this color map,  the theoretical Landau spectrum for the three electron pockets, e$_1$ (dotted line),  e$_2$(solid line) and $e_3$ (dash-dot line) are shown.}
\label{MapZoom}
\end{figure}

We have compared different sets of parameters in the model and opted for the set yielding the best agreement with experiment. These parameters yield an angular dependence for the electron lines, which still differ from the experimental results by a visible margin. This can be seen in Fig. \ref{MapZoom}, which is a zoom on the region in the vicinity of the trigonal axis. The superposition between the theoretical solid lines and the experimental color map is good but not perfect. Any attempt to repair this imperfection within the model used here led to a departure from the quite satisfactory agreement between theory and experiment on the subject of the angular variation of the hole lines.

The imperfect agreement between theory and experiment regarding the electron lines may point to some deeper physics associated with electron interaction. A recent theory\cite{abanin} suggests that Coulomb interaction in presence of strong mass anisotropy can lead to an inequality in the occupation of the three equivalent electron valleys.  Here, the relative agreement between theory and experiment puts an upper limit on any hypothetical inequality in the occupancy of the three equivalent electron valleys. Given the impressive agreement between theory and experiment in the case of holes, however, the relative disagreement in case of electrons (in particular for an arbitrarily-oriented magnetic field) may be speculatively attributed to a mild  ``spontaneous valley polarization''.

\section{The Zeeman splitting of holes}

A particularly important issue is the Zeeman splitting of the holes. When the field is along the trigonal axis, at low enough temperature, hole peaks are twin peaks, save for the one at the highest field at 9 T. This is a consequence of Zeeman splitting. The absence of splitting for the 9T peak associated with the $n=1$ Landau level can be naturally understood if one assumes that the Landau indexes of the two adjacent peaks differ by two. In other words, the  Zeeman energy creates a quasi-degeneracy between the $(n, +)$ and $(n\!-\!2, -)$ Landau levels, when the field is parallel to the trigonal axis. On the other hand, when the field is perpendicular to the trigonal axis, the Zeeman energy is vanishingly small.

These features have been known since Smith and co-workers\cite{smith}. From the position of the Zeeman-split peaks in the B $\|$ trigonal configuration, they concluded that $E_{z}/\hbar\omega_c$  is either 2.16 or 1.84. Indeed, the hole spectrum for B//trigonal does not allow to distinguish between these two possibilities. The Zeeman energy can be slightly larger or slightly lower than twice the cyclotron energy according to the way one indexes the twin peaks. Previous studies, focused on the spectrum in the vicinity of the trigonal axis, used either the former value \cite{bompadre,behnia2} or the latter\cite{alicea}. On the other hand, there is no ambiguity in indexing the Landau levels in the B $\bot$ trigonal configuration and, thus, the angular map allows to definitely settle this issue. The model used here assumes that $E_{z}/\hbar\omega_c=2.12$  and, as seen in Fig. \ref{trigbis} and Fig. \ref{trigbin}, successfully reproduces the complex experimentally-resolved angular dependence of the hole lines and in particular the angle at which they cross each other.

\begin{figure}
\resizebox{!}{0.3\textwidth}{\includegraphics[bb=0 0 506 309]{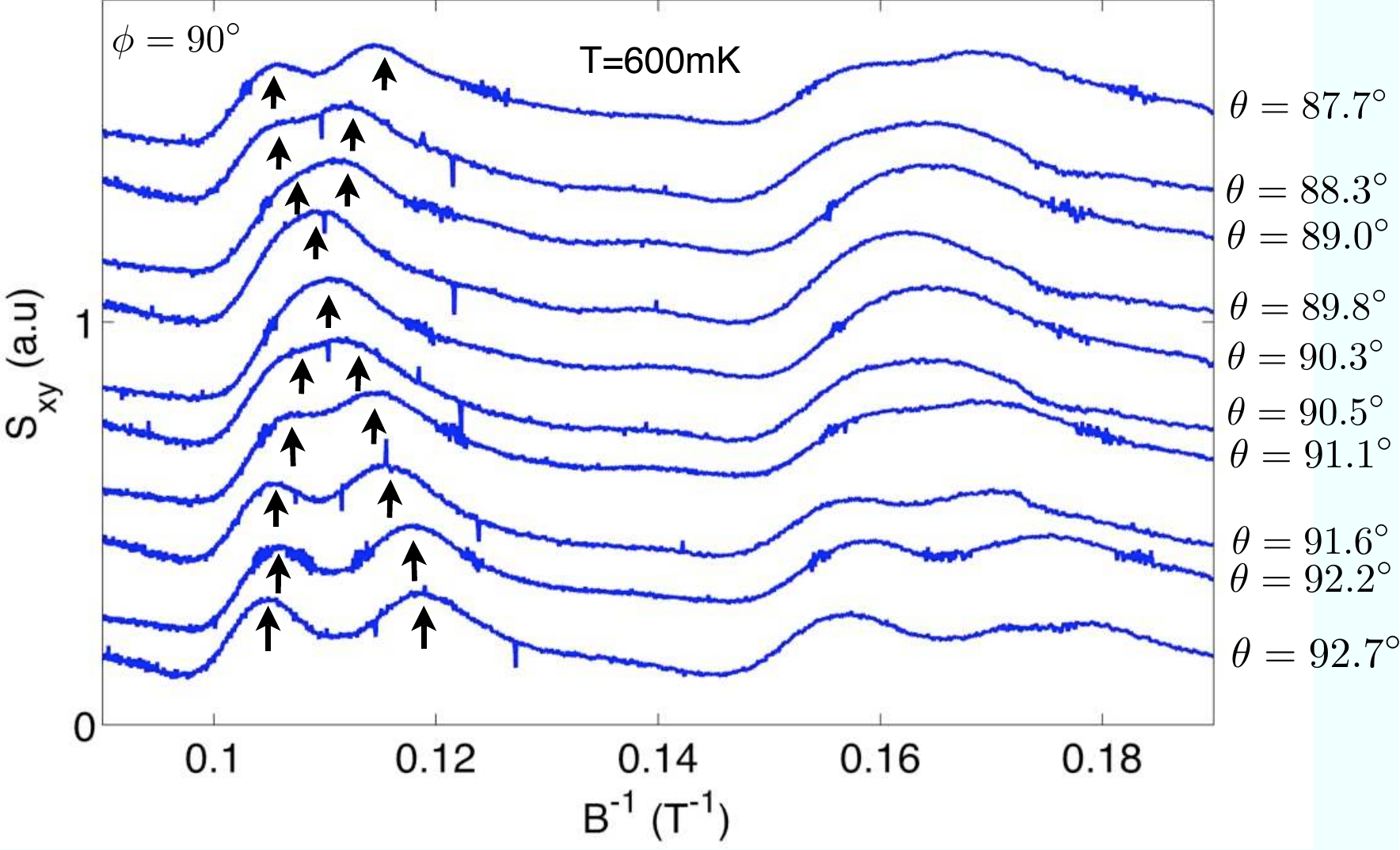}}\caption{Nernst voltage as a function of B$^{-1}$ for
$\phi=90^\circ$ as $\theta$ is scanned in the vicinity of 90$^\circ$, which correspond to the bisectrix direction.The curve are shifted for clarity. The black arrows indicate the position of two hole peaks with opposite spins($4^{+}$ and $4^{-}$). Near $\theta= 90^\circ$, experiment cannot resolve two distinct peaks. In other words, the Zeeman splitting becomes smaller than the width of a peak.}
\label{Holegfactor}
\end{figure}

When the field becomes perpendicular to the trigonal axis the splitting is below the threshold of experimental resolution even in the subkelvin temperature range and in a field of 9 T. This can be seen in Fig.\ref{Holegfactor}, which tracks the evolution of the hole Zeeman splitting in the vicinity of $\theta=90^\circ$. In other words, $E_{z}/\hbar\omega_c$ is lower than the experimental resolution in agreement with theory (See table II).

\section{Summary}

Our angle-dependent Nernst measurements map the Landau spectrum of bismuth up to 12 T. The experimental results are confronted with a theoretical model treating holes as quasi-particles with a parabolic dispersion and electrons as Dirac fermions. Within this model, we found a set of parameters, which produce theoretical results in  good agreement with the experimentally resolved spectrum in the whole solid angle. The model:
	\begin{enumerate}
		\item Fits almost the entire hole spectrum very well.
        \item Finds $E_{\rm Z}/\hbar \wc = 2.12$ for holes when the field is along the trigonal axis in agreement with the experimental value \cite{bompadre,behnia2}.
		\item Yields a satisfactory description of the level cross between $0_{\rm e}^+$ and $1_{\rm e}^-$ when the field rotates in the (Binary, Bisectrix). plane.
		\item Assumes a tilt angle for the electron pockets,  $6.2^\circ$, which is in agreement with previous low-field studies\cite{bhargava}.
		
 		\item Concludes that $1_{\rm h}^-$ and $1_{\rm e}^-$ meet at $B=9$T when the field is oriented along the trigonal axis in agreement with experiment.
	\end{enumerate}
On the other hand, the model is not entirely successful. In particular, the fine structure near the trigonal axis according to theory and experiment are somewhat different.

An important conclusion of this analysis is that when a field exceeding 2.5 T is applied along the bisectrix axis, all electrons reside in their
lowest ($j=0$) spin-polarized Landau level.

Finally, let us add that additional Nernst peaks, unexpected in the non-interacting picture, were resolved on top of this complex spectrum. The explanation of their origin is a subject of ongoing research.

\section{Acknowledgements}
We thank J. Alicea, L. Balents, G. Mikitik and Y. Sharlai for many interesting discussions. This work was supported in France by ANR as part of DELICE(ANR-08-BLAN-0121-02) and QUANTHERM(ANR-2010-INTB-401-01) projects and in Japan by the Institutional Program for Young Researcher Overseas Visits, Grant-in-Aid for Research Activity Start-up (No. 21840035) and Young Scientists (B) (No. 23740269) from Japan Society for the Promotion of Science.


\begin{thebibliography}{}
\bibitem{note1} This model is often called the Lax model. Here we will call it the Dirac model in order to highlight its link to other cases of  solid-state Dirac electrons.
\bibitem{note2} The original works by Cohen-Blount and Wolff initially included a term quadratic in $k$, $k^2/2m$. The quadratic term was omitted afterwards, since it appeared to be very small in bismuth. If this term is kept, it would be much more difficult to solve the problem in presence of magnetic field.
\end{thebibliography}
\end{document}